\documentclass[journal=nalefd,manuscript=letter,layout=onecolumn]{achemso}

\usepackage{amsmath}           
\usepackage{graphicx}          
\usepackage{float}

\author{Luca Sacchi}
\email{lsacchi@seas.harvard.edu}
\affiliation[Harvard University]
{Harvard John A. Paulson School of Engineering and Applied Sciences, Harvard University, Cambridge, Massachusetts, USA}
\alsoaffiliation[ETH]
{Department of Physics, ETH Zürich, Zürich, Switzerland}
\author{Alfonso Palmieri}
\affiliation[Harvard University]
{Harvard John A. Paulson School of Engineering and Applied Sciences, Harvard University, Cambridge, Massachusetts, USA}
\author{Vitthal Mishra}
\affiliation[INESC]
{Department of Physics, Instituto Superior Técnico, Universidade de Lisboa, Lisbon, Portugal}
\alsoaffiliation[Lisbon University]
{Instituto de Engenharia de Sistemas e Computadores---Microsistemas e Nanotecnologias (INESC MN), Lisbon, Portugal}
\author{Joon-Suh Park}
\affiliation[Harvard University]
{Harvard John A. Paulson School of Engineering and Applied Sciences, Harvard University, Cambridge, Massachusetts, USA}
\author{Marco Piccardo}
\affiliation[INESC]
{Department of Physics, Instituto Superior Técnico, Universidade de Lisboa, Lisbon, Portugal}
\alsoaffiliation[Lisbon University]
{Instituto de Engenharia de Sistemas e Computadores---Microsistemas e Nanotecnologias (INESC MN), Lisbon, Portugal}
\author{Federico Capasso}
\email{capasso@seas.harvard.edu}
\affiliation[Harvard University]
{Harvard John A. Paulson School of Engineering and Applied Sciences, Harvard University, Cambridge, Massachusetts, USA}

\title{Silica meta-optics: When high-performance does not need a high-index}

\begin{document}

    \begin{abstract}
        Metasurfaces---planar arrays of subwavelength nanostructures---are typically realized with high-index dielectrics, while low-index platforms are often dismissed for their weaker contrast. Here, we identify and experimentally verify regimes where a low-index platform (SiO$_2$) surpasses a high-index counterpart (TiO$_2$). We demonstrate that a low index suppresses higher-order Bloch modes, enabling the design of efficient devices with relaxed feature sizes. Low-index metasurfaces also offer two intrinsic advantages: a broad, well-behaved chromatic response without the need for explicit dispersion engineering, and strong tolerance to fabrication errors. We validate these features experimentally with silica metagratings, metalenses, and structured-light phase plates at $\lambda=632\ nm$. The metagratings reach $\geq$50\% absolute diffraction efficiency over a $200\ nm$ bandwidth, the metalenses deliver 75\% absolute diffraction efficiency with diffraction-limited performance, and the vortex phase plates achieve 80\% conversion efficiency at the design wavelength and 60\% with $100\ nm$ wavelength detuning. These results delineate conditions where low-index metasurfaces outperform high-index designs, suggesting a route to scalable, broadband, fabrication error-resilient flat optics.
    \end{abstract}

Recently, metasurface-based wavefront shaping has emerged as a major research focus \cite{yu_flat_2014,kuznetsov_roadmap_2024,lalanne_metalenses_2017,kamali_review_2018,dorrah_tunable_2022}. These thin structures can tailor the phase, amplitude, polarization, and dispersion of light, enabling advances in imaging \cite{chen_flat_2020,zhang_metasurfaces_2020,pahlevaninezhad_metasurface-based_2022,lenaerts_polychromatic_2025,cordaro_solving_2023,yang_freeform_2018}, sensing \cite{yesilkoy_ultrasensitive_2019,juliano_martins_metasurface-enhanced_2022}, polarization control \cite{rubin_matrix_2019,xie_generalized_2021,arbabi_full-stokes_2018,palmieri_bilayer_2023,palmieri_free_2025,dorrah_free-standing_2025}, optical communications \cite{oh_metasurfaces_2024}, atom trapping \cite{hsu_single-atom_2022}, and nonlinear optics \cite{wang_nonlinear_2018,tseng_vacuum_2022,andberger_terahertz_2024}. Metasurfaces are highly versatile due to the many geometric and material design degrees of freedom accessible via conventional nanofabrication processes. Key parameters include the meta-atom height, shape, center-to-center spacing ($U$), and material composition. Material selection is mainly guided by refractive index: higher index contrast is typically preferred since it allows full $(0,2\pi)$ phase coverage with fewer structural constraints, which is essential for many optical functionalities.

Various materials have been explored in metasurface designs \cite{yang_analysis_2017,choudhury_material_2018,yang_revisiting_2023,bayati_role_2019}. Silicon stands out for its high refractive index (Si, $n_\text{Si}\approx 3.4$) and established nanofabrication methods. Similarly, III–V semiconductors such as GaAs and InAs, with dielectric constants comparable to Si, are promising for efficient metasurfaces in the mid- and near-infrared \cite{chae_gaas_2024}. In the visible, transparent dielectrics like Titanium Dioxide (TiO$_2$, $n_\text{TiO$_2$} \approx 2.4$) and Gallium Nitride (GaN, $n_\text{GaN} \approx 2.3$) are attractive due to their low absorption \cite{khorasaninejad_metalenses_2016,chen_high-performance_2021}. Recently, pure silica-based metasurfaces (SiO$_2$, $n_\text{SiO$_2$}\approx1.45$) have gained interest despite the low-index contrast \cite{park_all-glass_2019,park_all-glass_2024,oliveira_high-aspect-ratio_2025}. Silica metasurfaces, despite their low-index contrast, offer unique advantages, like robustness under high-power laser illumination \cite{ray_substrate-engraved_2020,ray_large_2021,oliveira_high-aspect-ratio_2025} and full CMOS-process compatibility, making them a durable and scalable platform for advanced imaging, beam shaping, structured light, and AR/VR photonics.

The choice of metasurface material must consider intended optical functionality, range of operational wavelength and power, fabrication processes, scalability, and the design approach (forward vs. inverse). No single material is universally superior; instead, one should systematically weigh these factors to identify the optimal material for a given application.

\begin{figure*}[!ht]
  \centering
    \includegraphics[width=1\textwidth]{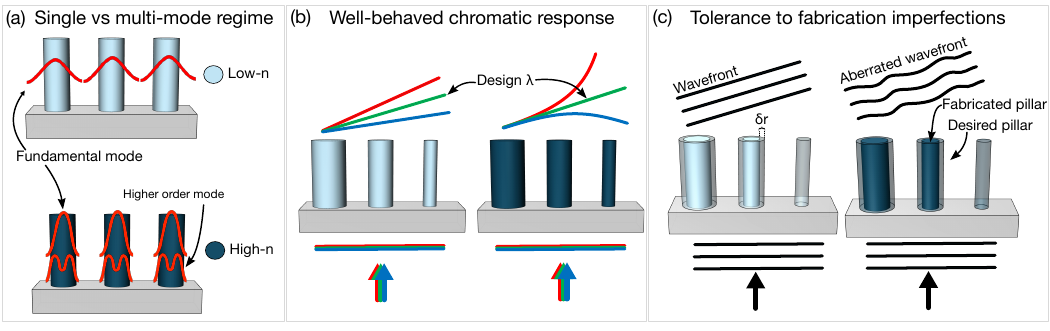}
    \caption{\textbf{Key advantages of low refractive index contrast metasurfaces.} \textbf{(a)} Single-mode and multi-mode Bloch representation in low-index and high-index contrast material nanopillars. When the unit cell is comparable in size to the operating wavelength, low-index pillars support fewer Bloch modes than high-index pillars, suppressing intermodal beating and preserving target phase gradients. \textbf{(b)} Broadband behavior of a low-index vs. high-index contrast material metasurface. Low material dispersion and modal dispersion of Bloch modes in low-index media pillars preserve a well-defined phase response over a wide spectral range.  \textbf{(c)} Robustness of optical functionality to fabrication error $\delta r$ in high-index vs low-index contrast materials. With lateral size errors $\delta r$ (shaded), high-index contrast metasurfaces lose the intended phase gradient, resulting in aberrated wavefronts, while low‑index contrast designs largely preserve the phase profile and functionality.}
    \label{fig:Main_Figure_1}
\end{figure*}

Here, we present a combined theoretical and experimental study to explore the forward design space of low-index contrast metasurfaces. Contrary to conventional expectations \cite{yang_freeform_2018,yang_revisiting_2023,lalanne_metalenses_2017}, we show that low-index metasurfaces can outperform high-index designs when the geometry is engineered so that each meta-atom supports only a single guided Bloch mode (Figure 1a). In this single-mode regime, intermodal beating is eliminated and the target phase gradient is preserved with high efficiency. Furthermore, a low-index contrast affords two key advantages: broad operational bandwidth (Figure 1b) and resilience to fabrication imperfections (Figure 1c). We validate these benefits through simulations on two representative low- and high-index material platforms, namely SiO$_2$ and TiO$_2$, and confirm them experimentally with silica meta-optics.

In a forward-design approach, a “library” of nanopillar optical responses (phase and amplitude) is first generated via numerical simulations (e.g., Difference Time-Domain method (FDTD) or Rigorous Coupled Wave Analysis (RCWA)) under a locally periodic approximation (LPA). This library maps each meta-atom geometry (height, lateral size, and unit cell pitch) to its complex transmission and reflection coefficients. Once the library is constructed, a target phase profile $\phi(x,y)$ is discretized into cells matching the chosen unit cell pitch, and each cell is assigned a pillar geometry that yields the required phase with maximal transmission. 

Because LPA assumes identical neighbors, a pillar’s actual response in a device (surrounded by dissimilar neighbors) may deviate from its library behavior. When the refractive index contrast is high (e.g., high-index pillars in air), relatively short heights suffice for $2\pi$ phase coverage, since the large index contrast increases the effective optical path length; approximately $\phi \approx 2\pi n_\text{eff} H/\lambda$, where $n_\text{eff}$ is the effective index of the unit cell and $H$ is the height of the pillar. However, high contrast also tends to excite multiple guided modes, which can interfere destructively. Understanding this modal landscape is therefore critical for optimizing performance.

For a metasurface to be maximally efficient, two conditions should be met. First, each unit cell must direct all power into the zeroth diffraction order, which by the grating equation \cite{yu_light_2011} requires the period to be smaller than the wavelength. A second and often overlooked constraint arises from the structural cut-off \cite{lalanne_optical_2006}: the unit cell size should be chosen so that only a single guided Bloch mode is supported for the entire range of possible fill factors. These are practical design rules for pixelated, forward-designed metasurfaces rather than fundamental limits. Inverse design approaches \cite{sell_large-angle_2017} reframe the problem by optimizing the entire metasurface as one scatterer, allowing the pillar geometry to vary continuously across the device and removing the need for a repeated unit cell. In this paradigm, multi-mode propagation, inter-pillar coupling, and higher diffraction orders are not drawbacks but additional design degrees of freedom to maximize a performance metric.

\begin{figure*}[!ht]
  \centering
    \includegraphics[width=0.84\textwidth]{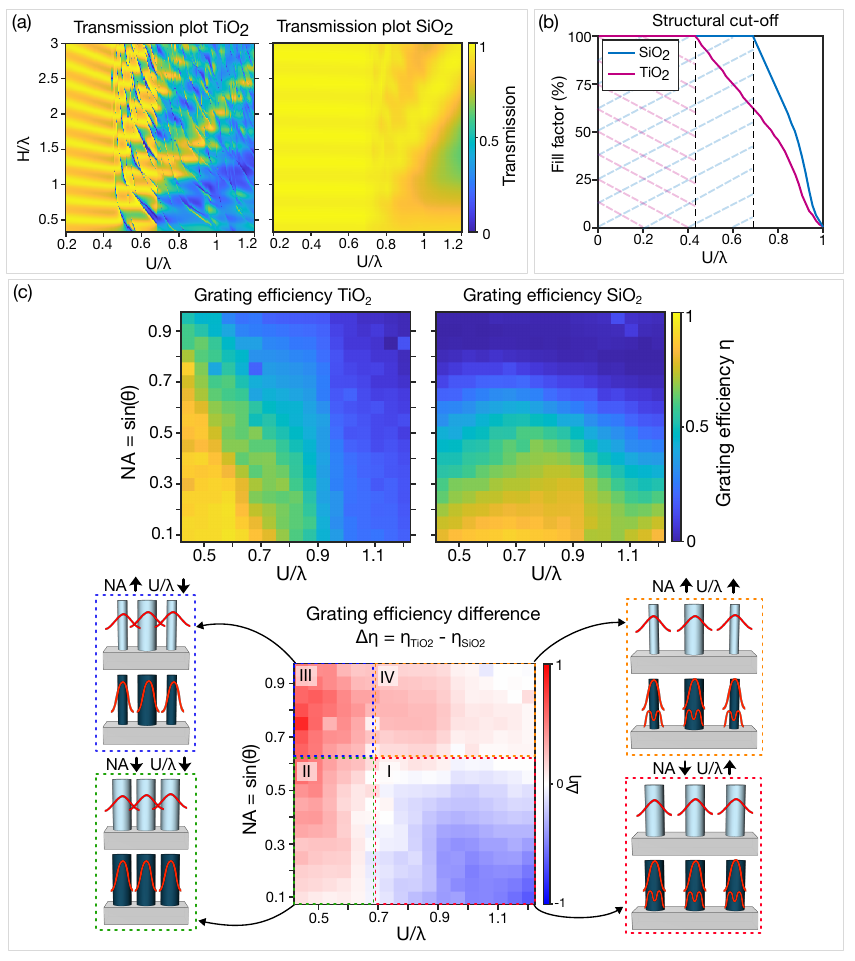}
    \caption{\textbf{Guided‑mode distribution and efficiency of high‑ vs. low‑index contrast metagratings.} \textbf{(a)} Transmission maps for TiO$_2$ and SiO$_2$ square pillars versus normalized pitch $U/\lambda$ and height $H/\lambda$; for each $U$, the pillar occupies 75\% of the unit‑cell area. \textbf{(b)} Onset of the first higher‑order Bloch mode versus $U/\lambda$ and fill factor (blue: SiO$_2$; magenta: TiO$_2$). Designs below the curve are single‑mode; above, multi-mode. Vertical dashed lines mark the structural cut‑off (the largest $U/\lambda$ for which all fill factors remain single‑mode), highlighting the wider single‑mode window of SiO$_2$. \textbf{(c)} Absolute diffraction efficiency $\eta$ (power in the target order divided by incident power on the metasurface) versus $U/\lambda$ and numerical aperture ($\text{NA}=\sin\theta$, with $\theta$ the deflection angle) for TiO$_2$ and SiO$_2$ gratings. The efficiency difference map $(\eta_{\text{TiO}_2}-\eta_{\text{SiO}_2})$ indicates regions where TiO$_2$ (red) or SiO$_2$ (blue) is superior. Insets schematically depict guided‑mode content and inter‑pillar coupling in each quadrant.}
  \label{fig:Main_Figure_2}
\end{figure*}

When a plane wave is normally incident on the metasurface, energy couples into guided Bloch-like modes in the pillars and is redistributed into the transmitted and reflected diffraction orders \cite{lalanne_optical_2006,yang_freeform_2018}. The number of supported Bloch modes increases with pillar width and index contrast, but decreases with wavelength. Figure 2a charts this interplay for a pillar occupying 75\% of the unit cell area, comparing SiO$_2$ and TiO$_2$. The simulated transmission amplitude is plotted versus unit cell size and pillar height (both normalized to the wavelength). Distinct regimes emerge, showing how geometry and scale determine the number of Bloch modes and the overall transmission. At low $U/\lambda$ (left side of each plot), the system operates in a single-mode regime characterized by slowly varying and high transmittance. In this regime, only the fundamental guided mode propagates, and higher-order modes are not excited. This single-mode region extends to considerably larger $U/\lambda$ for the low-index material, since weaker index contrast supports fewer guided modes (Supplementary Note 1 confirms this trend across different fill factors).

Figure 2b quantifies the structural cut-off for each material. The solid curves indicate the onset of the first higher-order mode as a function of normalized period $U/\lambda$ and fill factor. Pillars below each curve are single-mode, whereas those above support multiple modes. The vertical dashed line in each plot marks the structural cut-off: the largest $U/\lambda$ at which all fill factors remain single-mode. Clearly, SiO$_2$ provides a much broader single-mode design space than TiO$_2$, permitting larger normalized periods.

This observation provides a key design insight: because single-mode operation maximizes transmission, using a lower-index material allows larger unit cells (relative to $\lambda$) without sacrificing efficiency. This is important technologically, since fabrication is often limited by minimum feature sizes achievable with optical lithography---the dominant method for large-scale metasurface production \cite{she_large_2018,park_all-glass_2019,hu_cmos-compatible_2020,zhang_high-efficiency_2023,park_all-glass_2024}. In practice, shorter design wavelengths push structures toward smaller dimensions, where advanced lithographic techniques face cost, throughput, and defect challenges. By allowing larger features at a given wavelength, low-index platforms relax these fabrication constraints, ensuring compatibility with scalable, industry-standard processes while broadening the applicability of metasurfaces to visible and ultraviolet wavelengths.

We performed large-scale simulations of metagratings, sweeping $U/\lambda$ and deflection angle $\theta$, while keeping the nanopillar aspect ratio (height/width) below 12 for both materials. Details on the libraries used to simulate the gratings can be found in Supplementary Note 2. Analyses for additional aspect ratios are provided in Supplementary Note 3. Gratings are used here as a benchmark (Supplementary Note 4) because they are basic building blocks of diffractive optics \cite{chen_dispersion-engineered_2023}. The deflection angle of the simulated gratings is expressed relative to that of an ideal lens with numerical aperture NA (where $\text{NA} = \sin\theta$). 

Figure 2c (top panels) shows that TiO$_2$ gratings achieve high absolute efficiency ($\eta_{\text{TiO}_2}$) at small $U/\lambda$ and low NA, and they remain efficient even at high NA. SiO$_2$ gratings (top right) likewise exhibit high efficiency at small $U/\lambda$ and low NA, and importantly, they continue to perform well at larger $U/\lambda$. 

The difference map $\eta_{\text{TiO}_2} - \eta_{\text{SiO}_2}$ (Fig. 2c, bottom) delineates four quadrants of relative performance. In quadrant I (bottom-right, NA = 0.1–0.6 and $U/\lambda \ge 0.6$), SiO$_2$ visibly outperforms TiO$_2$ since the lower index contrast favors single-mode propagation. By contrast, the TiO$_2$ designs presented here are largely multimodal, with several Bloch modes propagating and lowering efficiency. In quadrant II (bottom-left, low NA and small $U/\lambda$), the hierarchy inverts. As $U/\lambda$ decreases, also TiO$_2$ enters the single-mode regime, the transmission increases, and its high index provides tight mode confinement. In contrast, SiO$_2$ (with weaker confinement) experiences stronger inter-pillar coupling, breaking the LPA and degrading the grating phase profile.

In short, there is a trade-off between unit cell size and inter-pillar coupling. Larger unit cells reduce coupling but push TiO$_2$ into multi-mode operation, which favors SiO$_2$. Smaller unit cells avoid multi-mode propagation, giving TiO$_2$ an edge due to superior mode confinement and minimal coupling. Thus, the choice of optimal material depends on which limitation---multi-mode excitation or inter-pillar coupling---dominates for a given design.

At high NA (quadrants III–IV), efficiency drops for both materials due to phase undersampling and strong inter‑pillar coupling, breaking the LPA. Rapid phase gradients are coarsely sampled---near or beyond the Nyquist limit ($U \sim \frac{\lambda_d}{2\text{NA}}$) \cite{kim_anti-aliased_2025}; reducing pitch mitigates undersampling but increases coupling. In this regime, TiO$_2$ outperforms SiO$_2$ owing to stronger mode confinement.

\begin{figure*}[!ht]
  \centering
    \includegraphics[width=0.83\textwidth]{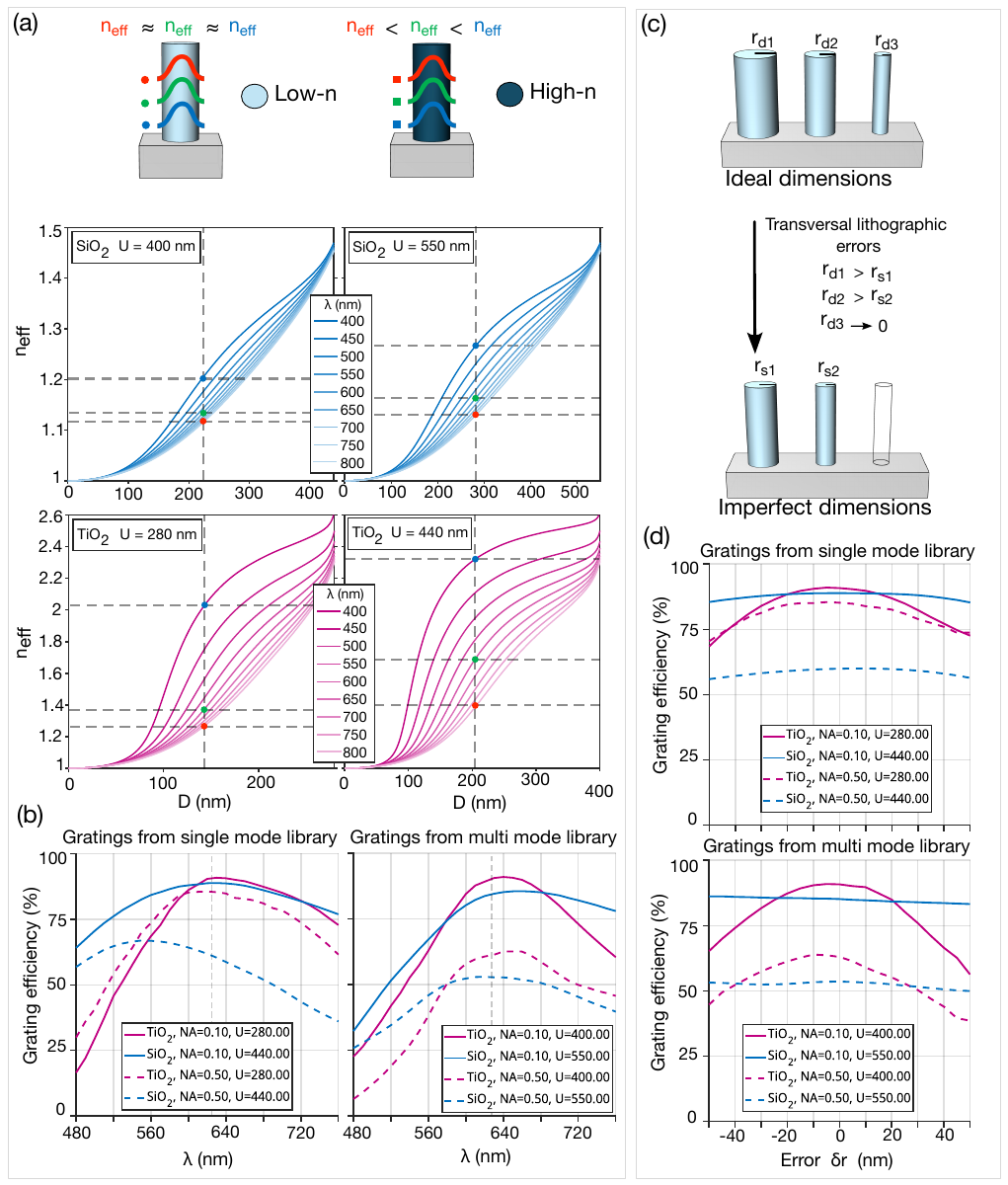}
    \caption{\textbf{Analysis of the dispersion behavior and fabrication tolerance of low-index contrast metasurfaces.} \textbf{(a)} Schematics and effective‑index ($n_\text{eff}$) dispersion of the fundamental Bloch mode for SiO$_2$ and TiO$_2$ pillar libraries when single-mode and multi-mode. For single-mode operation, we used pitches $U = 400\ nm$ for SiO$_2$ and $U = 280\ nm$ for TiO$_2$. For multi-mode behavior, we used $U = 550\ nm$ for SiO$_2$ and $U = 440\ nm$ for TiO$_2$. Blue, green, and red markers denote $\lambda=400$, $600$, and $800\ nm$, respectively. \textbf{(b)} Absolute diffraction efficiencies of simulated gratings when drawn from a single-mode library or a multi-mode library. The gray dashed line marks the design wavelength $\lambda_d=632\ nm$. Legend applies to both plots. \textbf{(c)} Schematic of a uniform lateral radius error $\delta r$ in the fabricated pillars. \textbf{(d)} Absolute diffraction efficiency of simulated gratings with $U=550\ nm$ and $U=280\ nm$ versus $\delta r$. Legend applies to both plots.}
  \label{fig:Main_Figure_3}
\end{figure*}

In addition to these geometric considerations, low-index contrast metasurfaces offer two additional intrinsic benefits: a well-behaved chromatic phase response around the design wavelength and greater tolerance to fabrication imperfections. Both advantages stem from the low material dispersion and, more importantly, the small refractive index contrast with the surroundings (in our case, air).

In the visible, low-index materials such as fused silica or polymer resins \cite{kim_scalable_2023} exhibit much lower material dispersion \cite{palik_handbook_1991} than high-index dielectrics like TiO$_2$ (Supplementary Note 5). Using a lower-index and less dispersive material inherently limits the effective index $n_{\text{eff}}(\lambda)$ of guided modes. Moreover, the weaker index contrast further reduces the waveguide contribution to the dispersion of $n_{\text{eff}}$. This effect is most pronounced for the fundamental Bloch mode, whose effective index lies between that of the cladding and the bulk pillar material ($n_{\text{clad}} \le n_{\text{eff}}(\lambda) \le n_{\text{bulk}}(\lambda)$). 

To investigate this phenomenon, we generated nanopillar libraries that span the full $(0,2\pi)$ phase range at a design wavelength $\lambda_d = 632\ nm$. For single-mode operation, we used pitches $U = 280\ nm$ for TiO$_2$ and $U = 400\ nm$ for SiO$_2$. For multi-mode behavior, we used $U = 440\ nm$ for TiO$_2$ and $U = 550\ nm$ for SiO$_2$. Supplementary Note 6 presents the full libraries. The libraries exhibit similar trends, and the onset and evolution of Bloch modes follow a similar qualitative behavior for both materials.

Figure 3a plots the effective index of the fundamental mode, $n_{\text{eff}}(D,\lambda)$, as a function of pillar diameter $D$ and wavelength $\lambda$. The curves $n_{\text{eff}}$ increase with $D$ and approach the bulk index $n_{\text{bulk}}(\lambda)$ as $D \to U$. Fixed $D$, the wavelength dependence of $n_{\text{eff}}$ is significantly weaker for SiO$_2$ than for TiO$_2$. This is due to the interplay of material and geometric dispersion: TiO$_2$’s higher material dispersion shifts $n_{\text{eff}}(\lambda)$ upward with decreasing $\lambda$. At the same time, its larger index contrast produces tighter field confinement, amplifying geometric dispersion and spreading the curves compared to SiO$_2$. 

In contrast, SiO$_2$ pillars show a nearly flat $n_{\text{eff}}(\lambda)$, especially in the single-mode regime where the fundamental mode dominates. These trends directly manifest in the metagrating absolute diffraction efficiencies (Figure 3b). When the gratings are built from libraries that are single-mode (Figure 3b left), both platforms are efficient near $\lambda_d$. At low NA, the SiO$_2$ grating is more broadband than the TiO$_2$ grating as a consequence of the low dispersion of the $n_{\text{eff}}$. At high NA, TiO$_2$ attains higher peak efficiency near $\lambda_d$ due to stronger mode confinement at smaller pitches (in accordance with Figure 2c), while SiO$_2$, although spectrally broader with a gentler roll-off, loses efficiency at $\lambda_d$ and its peak is blue-shifted. Even when the gratings are built from libraries that are multi-mode (Figure 3b right), and at both low and high NA, the SiO$_2$ gratings are more broadband and show a gentler efficiency roll-off than the TiO$_2$.

We next examined the effect of systematic pillar size errors, a common lithographic imperfection \cite{park_all-glass_2024}. Starting from ideal metagratings designed for a linear phase ramp (Figure 3c), we introduced a uniform radius offset $\delta r$ to every pillar: each pillar radius was set to $r_s = r_d + \delta r$, where $r_d$ is the design value. To maintain physical dimensions, if $r_s \le 0$ the pillar is removed, and if $r_s \ge U$ the cell is fully filled. We varied $\delta r$ from $-50\ nm$ to $+50\ nm$ and computed the absolute diffraction efficiency (Figure 3d) for both SiO$_2$ and TiO$_2$ at low deflection (NA = 0.1) and high deflection (NA = 0.5), for the two cases of a library fully single-mode or multi-mode, as in Figure 3b.

If fabricated pillar diameters deviate uniformly from the design, each meta-atom will impart a slightly incorrect phase. However, if the phase response of each pillar varies approximately linearly with its diameter ($\phi \approx \frac{2\pi}{\lambda} n_{\text{eff}}(D) H$ and $n_{\text{eff}} \propto D$) ---conditions that hold in the single-mode regime--- then a global size bias acts mainly as a uniform phase shift and does not severely degrade performance. This linear phase–diameter relationship is a good approximation for low-index pillars (Figure 3a), making them inherently more fault-tolerant.

In all the scenarios presented in Figure 3d, that is, at high deflection angle and low deflection angle for both when the grating is drawn from a single-mode library or multi-mode library, SiO$_2$ is markedly robust: the efficiency remains essentially constant across the entire bias range. For low-index contrast libraries, the pillar phase response is linear enough even when multiple modes are present, and a uniform size offset simply rescales the phase ramp, preserving the blazed profile and keeping higher diffraction orders suppressed. This gives a broad fabrication process window in which performance is maintained despite realistic size variations. By contrast, TiO$_2$ shows a pronounced efficiency degradation with increasing $|\delta r|$. At larger diameters, TiO$_2$ pillars start supporting higher-order modes, making the phase–diameter relation nonlinear and reducing transmitted amplitude. Consequently, a global size bias in a high-index device perturbs both the linear phase gradient and the local transmission, degrading the efficiency and redirecting power into undesired orders. In summary, SiO$_2$ also offers far greater tolerance to fabrication errors. As a result, any realistic fabrication offset quickly reverses the optimal choice between SiO$_2$ and TiO$_2$, where the former outperforms the latter. 

This study then highlights a practical advantage of low–index contrast platforms: high absolute efficiency combined with a wide process window that simplifies manufacturing, improves yield, and reduces sensitivity to process variations.

\begin{figure*}[!ht]
  \centering
    \includegraphics[width=0.84\textwidth]{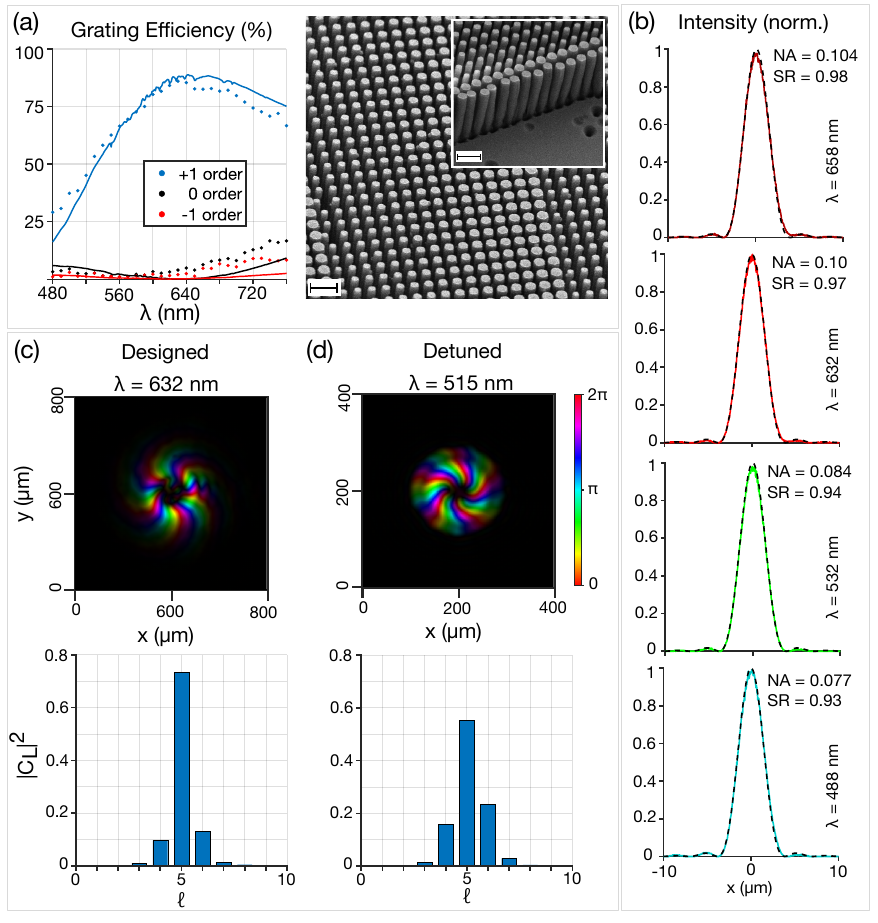}
    \caption{\textbf{Broadband measurements of silica metasurfaces.} \textbf{(a)} Measured absolute diffraction efficiency and Scanning Electron Microscopy (SEM) image of the metagrating. The scatter points represent experimental data, while the solid lines indicate simulated results. The grating consists of 16 nanopillars arranged in a $550\ nm$ unit cell, which produces a deflection angle of 4 degrees (NA = $\sin\theta$ = 0.07). The scale bar in both the SEM image and the inset corresponds to 1 $\mu m$. \textbf{(b)} Cut of the raw data Point Spread Function (PSF) of the metalens designed for NA = 0.1 at $\lambda_d=632\ nm$ and measured at various wavelengths. From top to bottom: $\lambda =658$, $632$, $532$, $488\ nm$. The lens shows diffraction-limited focusing for the measured wavelengths with Strehl ratios $\geq 0.8$. The effective-NA varies from 0.104 to 0.077 from $\lambda=658$ to $488$ nm, respectively. \textbf{(c)} Reconstructed phase of a measured orbital angular momentum (OAM) beam with a topological charge of $\ell = 5$ at $\lambda_d=632\ nm$ (upper plot) and modal decomposition spectrum on the azimuthal ($\ell$) modes of a Laguerre–Gaussian basis (lower plot). \textbf{(d)} Same as (c) but at a detuned wavelength of $515\ nm$.}
  \label{fig:Main_Figure_4}
\end{figure*}

To validate the broadband behavior of low-index metasurfaces, we fabricated several SiO$_2$ metasurfaces using a high-aspect-ratio etch process with a ruthenium hard mask \cite{oliveira_high-aspect-ratio_2025,okatani_fabrication_2023,mitchell_highly_2021} (Supplementary Note 7). The fabricated silica metasurfaces are assembled using a library with $U = 550\ nm$, considering an AR of 12 and a minimum feature size of $250\ nm$ for a height of nanopillars $H = 3000\ nm = 12\times 250\ nm$. The aspect ratio and unit cell chosen are consistent with all the simulation analyses in the paper; we chose a minimum feature size of $250\ nm$ due to the resolution constraint in the mask.

First, a diffraction grating (NA = 0.07, period $U=550\ nm$) was designed for $\lambda_d=632\ nm$ and measured. As shown in Figure 4a, the SiO$_2$ grating maintained an absolute diffraction efficiency above 50\% over a $200\ nm$ bandwidth around $\lambda_d$, in excellent agreement with simulation.

We also fabricated and characterized (see Supplementary Note 8 for experimental apparatus) a low-NA ($\text{NA}=0.1$) silica metalens with a hyperbolic phase profile $\phi(r) = -\frac{2\pi}{\lambda_d}(\sqrt{r^2+f_d^2}-f_d)$ at $\lambda_d=632\ nm$. For low NAs, the hyperbolic phase can be approximated by the parabolic phase profile $\phi(r) \approx \frac{\pi r^2}{\lambda_d f_d}$ (paraxial limit). Thanks to silica’s nearly achromatic $n_{\text{eff}}$ (Figure 3a), the focal length scales with wavelength as $f(\lambda) \approx (\lambda_d/\lambda)f_d$ (Supplementary Note 9), giving a wavelength-dependent $\text{NA}(\lambda)$. We computed the ideal diffraction-limited point spread functions (PSFs) for $\text{NA}(\lambda)$ and compared them to measurements. Figure 4b shows that at $\lambda = 658$, $632$, $532$, and $488\ nm$, the measured focal spot profiles (solid lines)  closely match the Airy pattern (dashed lines), and all the calculated Strehl ratios ($SR$) are above the diffraction limit ($SR\geq0.8$). The metalens not only remains diffraction-limited, but it is also efficient across the visible, achieving absolute diffraction efficiencies of 70\%, 74\%, 52\%, and 48\% at $658$, $632$, $532$, and $488\ nm$, respectively. For higher-NA hyperbolic lenses, the parabolic approximation breaks down and aberrations appear away from $\lambda_d$ (Supplementary Note 10).

As a further demonstration of broadband performance for structured light, we realized a SiO$_2$ vortex phase plate (topological charge $\ell=5$) designed for $\lambda_d = 632\ nm$. This device was also tested at a detuned wavelength of $515\ nm$. Unlike metalenses or gratings, which demand dispersion engineering for achromatic behavior, a vortex phase mask has a wavelength-independent phase profile and can operate achromatically provided each pillar’s phase response is achromatic (Supplementary Note 9). Using off-axis digital holography \cite{piccardo_broadband_2023}, we measured the output phase and decomposed the beam into Laguerre–Gaussian modes \cite{pinnell_modal_2020}. At $\lambda_d=632\ nm$, the device achieved 80\% relative conversion efficiency with high mode purity concentrated in the $\ell=5$ component (Figure 4c). At $\lambda=515\ nm$ (a 19\% detuning in wavelength), it retained 60\% conversion efficiency and 55\% purity in the $\ell=5$ mode, only about 20\% below the designed-wavelength performance (Figure 4d). Thus, although the ring diameter and accumulated phase scale with $\lambda$, the topological charge and dominant mode content are preserved with high efficiency across a broad band.

\begin{figure*}[h!]
  \centering
    \includegraphics[width=1\textwidth]{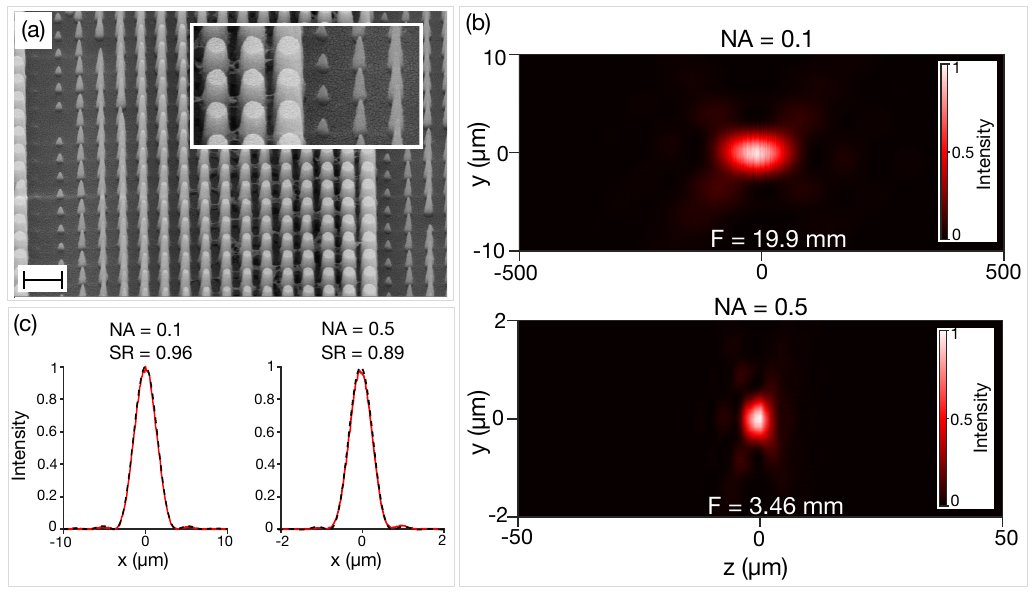}
    \caption{\textbf{Effect of fabrication error on metasurface performance.} \textbf{(a)} Scanning Electron Microscopy image of the metasurface with fabrication imperfection (scale bar 1 $\mu m$). The inset highlights the zoomed-in fabrication error. \textbf{(b)} Focusing profile for two silica lenses of NA = 0.1 (top) and NA = 0.5 (bottom) with fabrication imperfections. \textbf{(c)} Point spread function at design wavelength $\lambda_d = 632$ nm. Both the high NA lens and low NA lens are diffraction-limited and focused light with an absolute diffraction efficiency of 71\% and 43\%, despite the fabrication errors.}
  \label{fig:Main_Figure_5}
\end{figure*}

To verify the predicted fabrication tolerance, we fabricated two silica metalenses (NA = 0.1 and 0.5) for $\lambda_d=632\ nm$ with $U=550\ nm$, deliberately introducing fabrication errors. The SEM image in Figure 5a reveals prominent deviations: an overall negative bias (undersized diameter) along with notable sidewall tapering and partial loss of the smallest features. These imperfections are more severe than the idealized uniform bias in our simulations, yet the measured performance still reflects the robustness of silica. Both lenses produced diffraction-limited foci at $\lambda_d$ (Figure 5b,c). The primary effect of the defects was reduced transmission rather than wavefront distortion \cite{park_all-glass_2024}, yielding absolute diffraction efficiencies of 71\% for NA 0.1 and 43\% for NA 0.5. These measured efficiencies can be benchmarked against the simulated grating limits at $U=550\ nm$ (Figure 3d): at $\lambda_d=632\ nm$, the maximum diffraction efficiency (without error) is 85\% for NA 0.1 and 53\% for NA 0.5. Thus, despite substantial fabrication errors, the silica metalenses lost only a modest fraction of efficiency.

Our study challenges the notion that low-index dielectrics are not suitable for meta-optics. By constraining designs to the single-guided-mode regime, low-index (exemplified by fused silica) metalenses can achieve absolute diffraction efficiencies greater than 75\% at $632\ nm$ and gratings with absolute diffraction efficiencies greater than 50\% over a $200\ nm$ band, while offering three practical advantages.

First, low-index materials relax the structural cut-off constraints: the onset of multi-mode propagation is pushed to larger unit cells, expanding the design space and enabling performance that surpasses TiO$_2$ for up to moderate NA and moderate to high $U/\lambda$.

Second, low material and modal dispersion allow efficient broadband operation around the design wavelength without elaborate dispersion engineering. For example, we demonstrated a silica metagrating with $\geq$50\% efficiency over a $200\ nm$ bandwidth, and low-NA silica metalenses that maintained diffraction-limited, high-efficiency focusing from $488\ nm$ to $658\ nm$. We also showed that a silica vortex plate preserves high conversion efficiency and mode purity even $100\ nm$ away from its design wavelength.

Finally, low-index contrast confers exceptional fabrication tolerance: silica gratings and lenses exhibited nearly constant efficiency under $\pm50\ nm$ of lithographic deviation, a level of error that would severely degrade TiO$_2$ devices.

These findings establish low-index metasurfaces as a compelling alternative for applications that demand larger meta-atom dimensions (relative to the wavelength), broadband functionality, CMOS compatibility \cite{park_all-glass_2019,park_all-glass_2024}, scalability \cite{ray_large_2021}, or high laser-induced damage thresholds \cite{bonod_full-silica_2021,piccardo_trends_2025}. Silica, in particular, leverages mature semiconductor processing---standard optical lithography and etching---without requiring exotic materials or custom process steps. Combined with the relaxed geometric requirements for high-efficiency operation, this means silica metasurfaces can be seamlessly integrated into existing foundry workflows for wafer-scale manufacturing.

Our work focused on visible-wavelength metasurfaces designed via forward design and fabricated with electron-beam lithography. At very high NA, even low-index designs still suffer efficiency losses from undersampling and inter-pillar coupling. Inverse design (topology optimization) approaches \cite{dainese_shape_2024,lupoiu_multi-agentic_2025} could address these limitations and bring low-index performance to that of high-index platforms, by exploiting multiple guided modes and accounting for inter-pillar interactions to ensure they interfere constructively. By elucidating the modal physics, providing practical design guidelines, and experimentally validating device performance, this work broadens the palette of viable metasurface materials and lays the groundwork for robust, scalable, and broadband flat-optical systems for imaging, sensing, and holography.

\section*{Acknowledgments}
We thank D. Cassara for the assistance in taking SEM images and M. Yessenov, J. Lu and P. Ratra for insightful discussions. We also thank T. Tapani for assistance with the OAM characterization, J. Ribeiro for the compression algorithm of metasurface design, and H. Pires for access to the 515 nm laser. The simulation work was in part performed using Tidy3D software from Flexcompute. This work was performed in part at the Harvard University Center for Nanoscale Systems (CNS); a member of the National Nanotechnology Coordinated Infrastructure Network (NNCI), which is supported by the National Science Foundation under NSF award no. ECCS-2025158. L.S., A.P., J.P., and F.C. acknowledge funding support from the Air Force Office of Scientific Research under Award Number FA9550-21-1-0312. M.P. and V.M. acknowledge funding from the European Research Council (ERC StG) under the European Union’s Horizon Europe research and innovation program (Grant agreement No. 101161858).

\bibliography{references}

\providecommand{\latin}[1]{#1}
\makeatletter
\providecommand{\doi}
  {\begingroup\let\do\@makeother\dospecials
  \catcode`\{=1 \catcode`\}=2 \doi@aux}
\providecommand{\doi@aux}[1]{\endgroup\texttt{#1}}
\makeatother
\providecommand*\mcitethebibliography{\thebibliography}
\csname @ifundefined\endcsname{endmcitethebibliography}  {\let\endmcitethebibliography\endthebibliography}{}
\begin{mcitethebibliography}{55}
\providecommand*\natexlab[1]{#1}
\providecommand*\mciteSetBstSublistMode[1]{}
\providecommand*\mciteSetBstMaxWidthForm[2]{}
\providecommand*\mciteBstWouldAddEndPuncttrue
  {\def\EndOfBibitem{\unskip.}}
\providecommand*\mciteBstWouldAddEndPunctfalse
  {\let\EndOfBibitem\relax}
\providecommand*\mciteSetBstMidEndSepPunct[3]{}
\providecommand*\mciteSetBstSublistLabelBeginEnd[3]{}
\providecommand*\EndOfBibitem{}
\mciteSetBstSublistMode{f}
\mciteSetBstMaxWidthForm{subitem}{(\alph{mcitesubitemcount})}
\mciteSetBstSublistLabelBeginEnd
  {\mcitemaxwidthsubitemform\space}
  {\relax}
  {\relax}

\bibitem[Yu and Capasso()Yu, and Capasso]{yu_flat_2014}
Yu,~N.; Capasso,~F. Flat optics with designer metasurfaces. \emph{13}, 139--150\relax
\mciteBstWouldAddEndPuncttrue
\mciteSetBstMidEndSepPunct{\mcitedefaultmidpunct}
{\mcitedefaultendpunct}{\mcitedefaultseppunct}\relax
\EndOfBibitem
\bibitem[Kuznetsov \latin{et~al.}()Kuznetsov, Brongersma, Yao, Chen, Levy, Tsai, Zheludev, Faraon, Arbabi, Yu, Chanda, Crozier, Kildishev, Wang, Yang, Valentine, Genevet, Fan, Miller, Majumdar, Fröch, Brady, Heide, Veeraraghavan, Engheta, Alù, Polman, Atwater, Thureja, Paniagua-Dominguez, Ha, Barreda, Schuller, Staude, Grinblat, Kivshar, Peana, Yelin, Senichev, Shalaev, Saha, Boltasseva, Rho, Oh, Kim, Park, Devlin, and Pala]{kuznetsov_roadmap_2024}
Kuznetsov,~A.~I. \latin{et~al.}  Roadmap for Optical Metasurfaces. \emph{11}, 816--865\relax
\mciteBstWouldAddEndPuncttrue
\mciteSetBstMidEndSepPunct{\mcitedefaultmidpunct}
{\mcitedefaultendpunct}{\mcitedefaultseppunct}\relax
\EndOfBibitem
\bibitem[Lalanne and Chavel()Lalanne, and Chavel]{lalanne_metalenses_2017}
Lalanne,~P.; Chavel,~P. Metalenses at visible wavelengths: past, present, perspectives. \emph{11}, 1600295\relax
\mciteBstWouldAddEndPuncttrue
\mciteSetBstMidEndSepPunct{\mcitedefaultmidpunct}
{\mcitedefaultendpunct}{\mcitedefaultseppunct}\relax
\EndOfBibitem
\bibitem[Kamali \latin{et~al.}()Kamali, Arbabi, Arbabi, and Faraon]{kamali_review_2018}
Kamali,~S.~M.; Arbabi,~E.; Arbabi,~A.; Faraon,~A. A review of dielectric optical metasurfaces for wavefront control. \emph{7}, 1041--1068\relax
\mciteBstWouldAddEndPuncttrue
\mciteSetBstMidEndSepPunct{\mcitedefaultmidpunct}
{\mcitedefaultendpunct}{\mcitedefaultseppunct}\relax
\EndOfBibitem
\bibitem[Dorrah and Capasso()Dorrah, and Capasso]{dorrah_tunable_2022}
Dorrah,~A.~H.; Capasso,~F. Tunable structured light with flat optics. \emph{376}, eabi6860\relax
\mciteBstWouldAddEndPuncttrue
\mciteSetBstMidEndSepPunct{\mcitedefaultmidpunct}
{\mcitedefaultendpunct}{\mcitedefaultseppunct}\relax
\EndOfBibitem
\bibitem[Chen \latin{et~al.}()Chen, Zhu, and Capasso]{chen_flat_2020}
Chen,~W.~T.; Zhu,~A.~Y.; Capasso,~F. Flat optics with dispersion-engineered metasurfaces. \emph{5}, 604--620\relax
\mciteBstWouldAddEndPuncttrue
\mciteSetBstMidEndSepPunct{\mcitedefaultmidpunct}
{\mcitedefaultendpunct}{\mcitedefaultseppunct}\relax
\EndOfBibitem
\bibitem[Zhang \latin{et~al.}()Zhang, Wong, Zeng, Bi, Tai, Dholakia, and Olivo]{zhang_metasurfaces_2020}
Zhang,~S.; Wong,~C.~L.; Zeng,~S.; Bi,~R.; Tai,~K.; Dholakia,~K.; Olivo,~M. Metasurfaces for biomedical applications: imaging and sensing from a nanophotonics perspective. \emph{10}, 259--293\relax
\mciteBstWouldAddEndPuncttrue
\mciteSetBstMidEndSepPunct{\mcitedefaultmidpunct}
{\mcitedefaultendpunct}{\mcitedefaultseppunct}\relax
\EndOfBibitem
\bibitem[Pahlevaninezhad \latin{et~al.}()Pahlevaninezhad, Huang, Pahlevani, Bouma, Suter, Capasso, and Pahlevaninezhad]{pahlevaninezhad_metasurface-based_2022}
Pahlevaninezhad,~M.; Huang,~Y.-W.; Pahlevani,~M.; Bouma,~B.; Suter,~M.~J.; Capasso,~F.; Pahlevaninezhad,~H. Metasurface-based bijective illumination collection imaging provides high-resolution tomography in three dimensions. \emph{16}, 203--211\relax
\mciteBstWouldAddEndPuncttrue
\mciteSetBstMidEndSepPunct{\mcitedefaultmidpunct}
{\mcitedefaultendpunct}{\mcitedefaultseppunct}\relax
\EndOfBibitem
\bibitem[Lenaerts \latin{et~al.}()Lenaerts, Cassara, Chevalier, Park, Sacchi, Lim, Pestourie, Ginis, Meretska, and Capasso]{lenaerts_polychromatic_2025}
Lenaerts,~J.; Cassara,~D.; Chevalier,~P.; Park,~J.-S.; Sacchi,~L.; Lim,~S. W.~D.; Pestourie,~R.; Ginis,~V.; Meretska,~M.; Capasso,~F. Polychromatic metalens in the {NIR} for {CO}2 detection. Photonic and Phononic Properties of Engineered Nanostructures {XV}. p PC1337709, Backup Publisher: International Society for Optics and Photonics\relax
\mciteBstWouldAddEndPuncttrue
\mciteSetBstMidEndSepPunct{\mcitedefaultmidpunct}
{\mcitedefaultendpunct}{\mcitedefaultseppunct}\relax
\EndOfBibitem
\bibitem[Cordaro \latin{et~al.}()Cordaro, Edwards, Nikkhah, Alù, Engheta, and Polman]{cordaro_solving_2023}
Cordaro,~A.; Edwards,~B.; Nikkhah,~V.; Alù,~A.; Engheta,~N.; Polman,~A. Solving integral equations in free space with inverse-designed ultrathin optical metagratings. \emph{18}, 365--372\relax
\mciteBstWouldAddEndPuncttrue
\mciteSetBstMidEndSepPunct{\mcitedefaultmidpunct}
{\mcitedefaultendpunct}{\mcitedefaultseppunct}\relax
\EndOfBibitem
\bibitem[Yang \latin{et~al.}()Yang, Sell, and Fan]{yang_freeform_2018}
Yang,~J.; Sell,~D.; Fan,~J.~A. Freeform Metagratings Based on Complex Light Scattering Dynamics for Extreme, High Efficiency Beam Steering. \emph{530}, 1700302\relax
\mciteBstWouldAddEndPuncttrue
\mciteSetBstMidEndSepPunct{\mcitedefaultmidpunct}
{\mcitedefaultendpunct}{\mcitedefaultseppunct}\relax
\EndOfBibitem
\bibitem[Yesilkoy \latin{et~al.}()Yesilkoy, Arvelo, Jahani, Liu, Tittl, Cevher, Kivshar, and Altug]{yesilkoy_ultrasensitive_2019}
Yesilkoy,~F.; Arvelo,~E.~R.; Jahani,~Y.; Liu,~M.; Tittl,~A.; Cevher,~V.; Kivshar,~Y.; Altug,~H. Ultrasensitive hyperspectral imaging and biodetection enabled by dielectric metasurfaces. \emph{13}, 390--396\relax
\mciteBstWouldAddEndPuncttrue
\mciteSetBstMidEndSepPunct{\mcitedefaultmidpunct}
{\mcitedefaultendpunct}{\mcitedefaultseppunct}\relax
\EndOfBibitem
\bibitem[Juliano~Martins \latin{et~al.}()Juliano~Martins, Marinov, Youssef, Kyrou, Joubert, Colmagro, Gâté, Turbil, Coulon, Turover, Khadir, Giudici, Klitis, Sorel, and Genevet]{juliano_martins_metasurface-enhanced_2022}
Juliano~Martins,~R.; Marinov,~E.; Youssef,~M. A.~B.; Kyrou,~C.; Joubert,~M.; Colmagro,~C.; Gâté,~V.; Turbil,~C.; Coulon,~P.-M.; Turover,~D.; Khadir,~S.; Giudici,~M.; Klitis,~C.; Sorel,~M.; Genevet,~P. Metasurface-enhanced light detection and ranging technology. \emph{13}, 5724\relax
\mciteBstWouldAddEndPuncttrue
\mciteSetBstMidEndSepPunct{\mcitedefaultmidpunct}
{\mcitedefaultendpunct}{\mcitedefaultseppunct}\relax
\EndOfBibitem
\bibitem[Rubin \latin{et~al.}()Rubin, D’Aversa, Chevalier, Shi, Chen, and Capasso]{rubin_matrix_2019}
Rubin,~N.~A.; D’Aversa,~G.; Chevalier,~P.; Shi,~Z.; Chen,~W.~T.; Capasso,~F. Matrix Fourier optics enables a compact full-Stokes polarization camera. \emph{365}, eaax1839\relax
\mciteBstWouldAddEndPuncttrue
\mciteSetBstMidEndSepPunct{\mcitedefaultmidpunct}
{\mcitedefaultendpunct}{\mcitedefaultseppunct}\relax
\EndOfBibitem
\bibitem[Xie \latin{et~al.}()Xie, Pu, Jin, Xu, Guo, Li, Gao, Ma, and Luo]{xie_generalized_2021}
Xie,~X.; Pu,~M.; Jin,~J.; Xu,~M.; Guo,~Y.; Li,~X.; Gao,~P.; Ma,~X.; Luo,~X. Generalized Pancharatnam-Berry Phase in Rotationally Symmetric Meta-Atoms. \emph{126}, 183902\relax
\mciteBstWouldAddEndPuncttrue
\mciteSetBstMidEndSepPunct{\mcitedefaultmidpunct}
{\mcitedefaultendpunct}{\mcitedefaultseppunct}\relax
\EndOfBibitem
\bibitem[Arbabi \latin{et~al.}()Arbabi, Kamali, Arbabi, and Faraon]{arbabi_full-stokes_2018}
Arbabi,~E.; Kamali,~S.~M.; Arbabi,~A.; Faraon,~A. Full-Stokes Imaging Polarimetry Using Dielectric Metasurfaces. \emph{5}, 3132--3140, Publisher: American Chemical Society\relax
\mciteBstWouldAddEndPuncttrue
\mciteSetBstMidEndSepPunct{\mcitedefaultmidpunct}
{\mcitedefaultendpunct}{\mcitedefaultseppunct}\relax
\EndOfBibitem
\bibitem[Palmieri \latin{et~al.}()Palmieri, Dorrah, Oh, Dainese, and Capasso]{palmieri_bilayer_2023}
Palmieri,~A.; Dorrah,~A.~H.; Oh,~J.; Dainese,~P.; Capasso,~F. Bilayer Berry Phase Metasurfaces with Linear Polarization Bases. Optica Imaging Congress (3D, {COSI}, {DH}, {FLatOptics}, {IS}, {pcAOP}) (2023), paper {FM}1F.2. p FM1F.2\relax
\mciteBstWouldAddEndPuncttrue
\mciteSetBstMidEndSepPunct{\mcitedefaultmidpunct}
{\mcitedefaultendpunct}{\mcitedefaultseppunct}\relax
\EndOfBibitem
\bibitem[Palmieri \latin{et~al.}()Palmieri, Dorrah, Park, and Capasso]{palmieri_free_2025}
Palmieri,~A.; Dorrah,~A.; Park,~J.-S.; Capasso,~F. Free standing bilayer metasurfaces: a new approach to wavefront control. Optical Design Automation. p~29\relax
\mciteBstWouldAddEndPuncttrue
\mciteSetBstMidEndSepPunct{\mcitedefaultmidpunct}
{\mcitedefaultendpunct}{\mcitedefaultseppunct}\relax
\EndOfBibitem
\bibitem[Dorrah \latin{et~al.}()Dorrah, Park, Palmieri, and Capasso]{dorrah_free-standing_2025}
Dorrah,~A.~H.; Park,~J.-S.; Palmieri,~A.; Capasso,~F. Free-standing bilayer metasurfaces in the visible. \emph{16}, 3126\relax
\mciteBstWouldAddEndPuncttrue
\mciteSetBstMidEndSepPunct{\mcitedefaultmidpunct}
{\mcitedefaultendpunct}{\mcitedefaultseppunct}\relax
\EndOfBibitem
\bibitem[Oh \latin{et~al.}()Oh, Yang, Marra, Dorrah, Palmieri, Dainese, and Capasso]{oh_metasurfaces_2024}
Oh,~J.; Yang,~J.; Marra,~L.; Dorrah,~A.~H.; Palmieri,~A.; Dainese,~P.; Capasso,~F. Metasurfaces for Free-Space Coupling to Multicore Fibers. \emph{42}, 2385--2396\relax
\mciteBstWouldAddEndPuncttrue
\mciteSetBstMidEndSepPunct{\mcitedefaultmidpunct}
{\mcitedefaultendpunct}{\mcitedefaultseppunct}\relax
\EndOfBibitem
\bibitem[Hsu \latin{et~al.}()Hsu, Zhu, Thiele, Brown, Papp, Agrawal, and Regal]{hsu_single-atom_2022}
Hsu,~T.-W.; Zhu,~W.; Thiele,~T.; Brown,~M.~O.; Papp,~S.~B.; Agrawal,~A.; Regal,~C.~A. Single-Atom Trapping in a Metasurface-Lens Optical Tweezer. \emph{3}, 030316\relax
\mciteBstWouldAddEndPuncttrue
\mciteSetBstMidEndSepPunct{\mcitedefaultmidpunct}
{\mcitedefaultendpunct}{\mcitedefaultseppunct}\relax
\EndOfBibitem
\bibitem[Wang \latin{et~al.}()Wang, Kruk, Koshelev, Kravchenko, Luther-Davies, and Kivshar]{wang_nonlinear_2018}
Wang,~L.; Kruk,~S.; Koshelev,~K.; Kravchenko,~I.; Luther-Davies,~B.; Kivshar,~Y. Nonlinear Wavefront Control with All-Dielectric Metasurfaces. \emph{18}, 3978--3984\relax
\mciteBstWouldAddEndPuncttrue
\mciteSetBstMidEndSepPunct{\mcitedefaultmidpunct}
{\mcitedefaultendpunct}{\mcitedefaultseppunct}\relax
\EndOfBibitem
\bibitem[Tseng \latin{et~al.}()Tseng, Semmlinger, Zhang, Arndt, Huang, Yang, Kuo, Su, Chen, Chu, Cerjan, Tsai, Nordlander, and Halas]{tseng_vacuum_2022}
Tseng,~M.~L.; Semmlinger,~M.; Zhang,~M.; Arndt,~C.; Huang,~T.-T.; Yang,~J.; Kuo,~H.~Y.; Su,~V.-C.; Chen,~M.~K.; Chu,~C.~H.; Cerjan,~B.; Tsai,~D.~P.; Nordlander,~P.; Halas,~N.~J. Vacuum ultraviolet nonlinear metalens. \emph{8}, eabn5644\relax
\mciteBstWouldAddEndPuncttrue
\mciteSetBstMidEndSepPunct{\mcitedefaultmidpunct}
{\mcitedefaultendpunct}{\mcitedefaultseppunct}\relax
\EndOfBibitem
\bibitem[Andberger \latin{et~al.}()Andberger, Graziotto, Sacchi, Beck, Scalari, and Faist]{andberger_terahertz_2024}
Andberger,~J.; Graziotto,~L.; Sacchi,~L.; Beck,~M.; Scalari,~G.; Faist,~J. Terahertz chiral subwavelength cavities breaking time-reversal symmetry via ultrastrong light-matter interaction. \emph{109}, L161302\relax
\mciteBstWouldAddEndPuncttrue
\mciteSetBstMidEndSepPunct{\mcitedefaultmidpunct}
{\mcitedefaultendpunct}{\mcitedefaultseppunct}\relax
\EndOfBibitem
\bibitem[Yang and Fan()Yang, and Fan]{yang_analysis_2017}
Yang,~J.; Fan,~J.~A. Analysis of material selection on dielectric metasurface performance. \emph{25}, 23899\relax
\mciteBstWouldAddEndPuncttrue
\mciteSetBstMidEndSepPunct{\mcitedefaultmidpunct}
{\mcitedefaultendpunct}{\mcitedefaultseppunct}\relax
\EndOfBibitem
\bibitem[Choudhury \latin{et~al.}()Choudhury, Wang, Chaudhuri, {DeVault}, Kildishev, Boltasseva, and Shalaev]{choudhury_material_2018}
Choudhury,~S.~M.; Wang,~D.; Chaudhuri,~K.; {DeVault},~C.; Kildishev,~A.~V.; Boltasseva,~A.; Shalaev,~V.~M. Material platforms for optical metasurfaces. \emph{7}, 959--987\relax
\mciteBstWouldAddEndPuncttrue
\mciteSetBstMidEndSepPunct{\mcitedefaultmidpunct}
{\mcitedefaultendpunct}{\mcitedefaultseppunct}\relax
\EndOfBibitem
\bibitem[Yang \latin{et~al.}()Yang, Kang, Jung, Seong, Jeon, Kim, Oh, Park, Kim, and Rho]{yang_revisiting_2023}
Yang,~Y.; Kang,~H.; Jung,~C.; Seong,~J.; Jeon,~N.; Kim,~J.; Oh,~D.~K.; Park,~J.; Kim,~H.; Rho,~J. Revisiting Optical Material Platforms for Efficient Linear and Nonlinear Dielectric Metasurfaces in the Ultraviolet, Visible, and Infrared. \emph{10}, 307--321\relax
\mciteBstWouldAddEndPuncttrue
\mciteSetBstMidEndSepPunct{\mcitedefaultmidpunct}
{\mcitedefaultendpunct}{\mcitedefaultseppunct}\relax
\EndOfBibitem
\bibitem[Bayati \latin{et~al.}()Bayati, Zhan, Colburn, Zhelyeznyakov, and Majumdar]{bayati_role_2019}
Bayati,~E.; Zhan,~A.; Colburn,~S.; Zhelyeznyakov,~M.~V.; Majumdar,~A. Role of refractive index in metalens performance. \emph{58}, 1460\relax
\mciteBstWouldAddEndPuncttrue
\mciteSetBstMidEndSepPunct{\mcitedefaultmidpunct}
{\mcitedefaultendpunct}{\mcitedefaultseppunct}\relax
\EndOfBibitem
\bibitem[Chae \latin{et~al.}()Chae, Shrewsbury, Ahsan, Povinelli, and Kapadia]{chae_gaas_2024}
Chae,~H.~U.; Shrewsbury,~B.; Ahsan,~R.; Povinelli,~M.~L.; Kapadia,~R. {GaAs} Mid-{IR} Electrically Tunable Metasurfaces. \emph{24}, 2581--2588\relax
\mciteBstWouldAddEndPuncttrue
\mciteSetBstMidEndSepPunct{\mcitedefaultmidpunct}
{\mcitedefaultendpunct}{\mcitedefaultseppunct}\relax
\EndOfBibitem
\bibitem[Khorasaninejad \latin{et~al.}()Khorasaninejad, Chen, Devlin, Oh, Zhu, and Capasso]{khorasaninejad_metalenses_2016}
Khorasaninejad,~M.; Chen,~W.~T.; Devlin,~R.~C.; Oh,~J.; Zhu,~A.~Y.; Capasso,~F. Metalenses at visible wavelengths: Diffraction-limited focusing and subwavelength resolution imaging. \emph{352}, 1190--1194\relax
\mciteBstWouldAddEndPuncttrue
\mciteSetBstMidEndSepPunct{\mcitedefaultmidpunct}
{\mcitedefaultendpunct}{\mcitedefaultseppunct}\relax
\EndOfBibitem
\bibitem[Chen \latin{et~al.}()Chen, Chou, Su, Kuan, and Lin]{chen_high-performance_2021}
Chen,~M.-H.; Chou,~W.-N.; Su,~V.-C.; Kuan,~C.-H.; Lin,~H.~Y. High-performance gallium nitride dielectric metalenses for imaging in the visible. \emph{11}, 6500\relax
\mciteBstWouldAddEndPuncttrue
\mciteSetBstMidEndSepPunct{\mcitedefaultmidpunct}
{\mcitedefaultendpunct}{\mcitedefaultseppunct}\relax
\EndOfBibitem
\bibitem[Park \latin{et~al.}()Park, Zhang, She, Chen, Lin, Yousef, Cheng, and Capasso]{park_all-glass_2019}
Park,~J.-S.; Zhang,~S.; She,~A.; Chen,~W.~T.; Lin,~P.; Yousef,~K. M.~A.; Cheng,~J.-X.; Capasso,~F. All-Glass, Large Metalens at Visible Wavelength Using Deep-Ultraviolet Projection Lithography. \emph{19}, 8673--8682\relax
\mciteBstWouldAddEndPuncttrue
\mciteSetBstMidEndSepPunct{\mcitedefaultmidpunct}
{\mcitedefaultendpunct}{\mcitedefaultseppunct}\relax
\EndOfBibitem
\bibitem[Park \latin{et~al.}()Park, Lim, Amirzhan, Kang, Karrfalt, Kim, Leger, Urbas, Ossiander, Li, and Capasso]{park_all-glass_2024}
Park,~J.-S.; Lim,~S. W.~D.; Amirzhan,~A.; Kang,~H.; Karrfalt,~K.; Kim,~D.; Leger,~J.; Urbas,~A.; Ossiander,~M.; Li,~Z.; Capasso,~F. All-Glass 100 mm Diameter Visible Metalens for Imaging the Cosmos. \emph{18}, 3187--3198\relax
\mciteBstWouldAddEndPuncttrue
\mciteSetBstMidEndSepPunct{\mcitedefaultmidpunct}
{\mcitedefaultendpunct}{\mcitedefaultseppunct}\relax
\EndOfBibitem
\bibitem[Oliveira \latin{et~al.}()Oliveira, Claveria, Araujo, Estrela, Gonçalves, Nunes, Meirinho, Fajardo, and Piccardo]{oliveira_high-aspect-ratio_2025}
Oliveira,~B.; Claveria,~P. S.~M.; Araujo,~P. D.~R.; Estrela,~P.; Gonçalves,~I.; Nunes,~M. I.~S.; Meirinho,~R.; Fajardo,~M.; Piccardo,~M. High-aspect-ratio, ultratall silica meta-optics for high-intensity structured light. \emph{12}, 713\relax
\mciteBstWouldAddEndPuncttrue
\mciteSetBstMidEndSepPunct{\mcitedefaultmidpunct}
{\mcitedefaultendpunct}{\mcitedefaultseppunct}\relax
\EndOfBibitem
\bibitem[Ray \latin{et~al.}()Ray, Yoo, Nguyen, Johnson, Elhadj, Baxamusa, and Feigenbaum]{ray_substrate-engraved_2020}
Ray,~N.~J.; Yoo,~J.-H.; Nguyen,~H.~T.; Johnson,~M.~A.; Elhadj,~S.; Baxamusa,~S.~H.; Feigenbaum,~E. Substrate-engraved antireflective nanostructured surfaces for high-power laser applications. \emph{7}, 518\relax
\mciteBstWouldAddEndPuncttrue
\mciteSetBstMidEndSepPunct{\mcitedefaultmidpunct}
{\mcitedefaultendpunct}{\mcitedefaultseppunct}\relax
\EndOfBibitem
\bibitem[Ray \latin{et~al.}()Ray, Yoo, Nguyen, and Feigenbaum]{ray_large_2021}
Ray,~N.~J.; Yoo,~J.-H.; Nguyen,~H.~T.; Feigenbaum,~E. Large aperture and durable glass-engraved optical metasurfaces using nanoparticle etching masks: prospects and future directions. \emph{3}, 032004\relax
\mciteBstWouldAddEndPuncttrue
\mciteSetBstMidEndSepPunct{\mcitedefaultmidpunct}
{\mcitedefaultendpunct}{\mcitedefaultseppunct}\relax
\EndOfBibitem
\bibitem[Yu \latin{et~al.}()Yu, Genevet, Kats, Aieta, Tetienne, Capasso, and Gaburro]{yu_light_2011}
Yu,~N.; Genevet,~P.; Kats,~M.~A.; Aieta,~F.; Tetienne,~J.-P.; Capasso,~F.; Gaburro,~Z. Light Propagation with Phase Discontinuities: Generalized Laws of Reflection and Refraction. \emph{334}, 333--337\relax
\mciteBstWouldAddEndPuncttrue
\mciteSetBstMidEndSepPunct{\mcitedefaultmidpunct}
{\mcitedefaultendpunct}{\mcitedefaultseppunct}\relax
\EndOfBibitem
\bibitem[Lalanne \latin{et~al.}()Lalanne, Hugonin, and Chavel]{lalanne_optical_2006}
Lalanne,~P.; Hugonin,~J.; Chavel,~P. Optical properties of deep lamellar Gratings: A coupled Bloch-mode insight. \emph{24}, 2442--2449\relax
\mciteBstWouldAddEndPuncttrue
\mciteSetBstMidEndSepPunct{\mcitedefaultmidpunct}
{\mcitedefaultendpunct}{\mcitedefaultseppunct}\relax
\EndOfBibitem
\bibitem[Sell \latin{et~al.}()Sell, Yang, Doshay, Yang, and Fan]{sell_large-angle_2017}
Sell,~D.; Yang,~J.; Doshay,~S.; Yang,~R.; Fan,~J.~A. Large-Angle, Multifunctional Metagratings Based on Freeform Multimode Geometries. \emph{17}, 3752--3757\relax
\mciteBstWouldAddEndPuncttrue
\mciteSetBstMidEndSepPunct{\mcitedefaultmidpunct}
{\mcitedefaultendpunct}{\mcitedefaultseppunct}\relax
\EndOfBibitem
\bibitem[She \latin{et~al.}()She, Zhang, Shian, Clarke, and Capasso]{she_large_2018}
She,~A.; Zhang,~S.; Shian,~S.; Clarke,~D.~R.; Capasso,~F. Large area metalenses: design, characterization, and mass manufacturing. \emph{26}, 1573\relax
\mciteBstWouldAddEndPuncttrue
\mciteSetBstMidEndSepPunct{\mcitedefaultmidpunct}
{\mcitedefaultendpunct}{\mcitedefaultseppunct}\relax
\EndOfBibitem
\bibitem[Hu \latin{et~al.}()Hu, Zhong, Li, Dong, Xu, Fu, Li, Bliznetsov, Zhou, Lai, Lin, Zhu, and Singh]{hu_cmos-compatible_2020}
Hu,~T.; Zhong,~Q.; Li,~N.; Dong,~Y.; Xu,~Z.; Fu,~Y.~H.; Li,~D.; Bliznetsov,~V.; Zhou,~Y.; Lai,~K.~H.; Lin,~Q.; Zhu,~S.; Singh,~N. {CMOS}-compatible a-Si metalenses on a 12-inch glass wafer for fingerprint imaging. \emph{9}, 823--830\relax
\mciteBstWouldAddEndPuncttrue
\mciteSetBstMidEndSepPunct{\mcitedefaultmidpunct}
{\mcitedefaultendpunct}{\mcitedefaultseppunct}\relax
\EndOfBibitem
\bibitem[Zhang \latin{et~al.}()Zhang, Chang, Chen, Ding, Rahman, Duan, Stephen, and Ni]{zhang_high-efficiency_2023}
Zhang,~L.; Chang,~S.; Chen,~X.; Ding,~Y.; Rahman,~M.~T.; Duan,~Y.; Stephen,~M.; Ni,~X. High-Efficiency, 80 mm Aperture Metalens Telescope. \emph{23}, 51--57\relax
\mciteBstWouldAddEndPuncttrue
\mciteSetBstMidEndSepPunct{\mcitedefaultmidpunct}
{\mcitedefaultendpunct}{\mcitedefaultseppunct}\relax
\EndOfBibitem
\bibitem[Chen \latin{et~al.}()Chen, Park, Marchioni, Millay, Yousef, and Capasso]{chen_dispersion-engineered_2023}
Chen,~W.~T.; Park,~J.-S.; Marchioni,~J.; Millay,~S.; Yousef,~K. M.~A.; Capasso,~F. Dispersion-engineered metasurfaces reaching broadband 90\% relative diffraction efficiency. \emph{14}, 2544\relax
\mciteBstWouldAddEndPuncttrue
\mciteSetBstMidEndSepPunct{\mcitedefaultmidpunct}
{\mcitedefaultendpunct}{\mcitedefaultseppunct}\relax
\EndOfBibitem
\bibitem[Kim \latin{et~al.}()Kim, Kim, Kim, Jeong, and Rho]{kim_anti-aliased_2025}
Kim,~S.; Kim,~J.; Kim,~K.; Jeong,~M.; Rho,~J. Anti-aliased metasurfaces beyond the Nyquist limit. \emph{16}, 411\relax
\mciteBstWouldAddEndPuncttrue
\mciteSetBstMidEndSepPunct{\mcitedefaultmidpunct}
{\mcitedefaultendpunct}{\mcitedefaultseppunct}\relax
\EndOfBibitem
\bibitem[Kim \latin{et~al.}()Kim, Seong, Kim, Lee, Kim, Kim, Moon, Oh, Yang, Park, Jang, Kim, Jeong, Park, Choi, Jeon, Lee, Yoon, Park, Lee, Lee, and Rho]{kim_scalable_2023}
Kim,~J. \latin{et~al.}  Scalable manufacturing of high-index atomic layer–polymer hybrid metasurfaces for metaphotonics in the visible. \emph{22}, 474--481\relax
\mciteBstWouldAddEndPuncttrue
\mciteSetBstMidEndSepPunct{\mcitedefaultmidpunct}
{\mcitedefaultendpunct}{\mcitedefaultseppunct}\relax
\EndOfBibitem
\bibitem[pal()]{palik_handbook_1991}
Palik,~E.~D., Ed. \emph{Handbook of optical constants of solids {II}}; Academic Press\relax
\mciteBstWouldAddEndPuncttrue
\mciteSetBstMidEndSepPunct{\mcitedefaultmidpunct}
{\mcitedefaultendpunct}{\mcitedefaultseppunct}\relax
\EndOfBibitem
\bibitem[Okatani \latin{et~al.}()Okatani, Naito, and Kanamori]{okatani_fabrication_2023}
Okatani,~T.; Naito,~Y.; Kanamori,~Y. Fabrication of high-aspect-ratio {SiO}$_{\textrm{2}}$ nanopillars by Si thermal oxidation for metalenses in the visible region. \emph{62}, SG1034\relax
\mciteBstWouldAddEndPuncttrue
\mciteSetBstMidEndSepPunct{\mcitedefaultmidpunct}
{\mcitedefaultendpunct}{\mcitedefaultseppunct}\relax
\EndOfBibitem
\bibitem[Mitchell \latin{et~al.}()Mitchell, Thibeault, John, and Reynolds]{mitchell_highly_2021}
Mitchell,~W.~J.; Thibeault,~B.~J.; John,~D.~D.; Reynolds,~T.~E. Highly selective and vertical etch of silicon dioxide using ruthenium films as an etch mask. \emph{39}, 043204\relax
\mciteBstWouldAddEndPuncttrue
\mciteSetBstMidEndSepPunct{\mcitedefaultmidpunct}
{\mcitedefaultendpunct}{\mcitedefaultseppunct}\relax
\EndOfBibitem
\bibitem[Piccardo \latin{et~al.}()Piccardo, De~Oliveira, Policht, Russo, Ardini, Corti, Valentini, Vieira, Manzoni, Cerullo, and Ambrosio]{piccardo_broadband_2023}
Piccardo,~M.; De~Oliveira,~M.; Policht,~V.~R.; Russo,~M.; Ardini,~B.; Corti,~M.; Valentini,~G.; Vieira,~J.; Manzoni,~C.; Cerullo,~G.; Ambrosio,~A. Broadband control of topological–spectral correlations in space–time beams. \emph{17}, 822--828\relax
\mciteBstWouldAddEndPuncttrue
\mciteSetBstMidEndSepPunct{\mcitedefaultmidpunct}
{\mcitedefaultendpunct}{\mcitedefaultseppunct}\relax
\EndOfBibitem
\bibitem[Pinnell \latin{et~al.}()Pinnell, Nape, Sephton, Cox, Rodríguez-Fajardo, and Forbes]{pinnell_modal_2020}
Pinnell,~J.; Nape,~I.; Sephton,~B.; Cox,~M.~A.; Rodríguez-Fajardo,~V.; Forbes,~A. Modal analysis of structured light with spatial light modulators: a practical tutorial. \emph{37}, C146\relax
\mciteBstWouldAddEndPuncttrue
\mciteSetBstMidEndSepPunct{\mcitedefaultmidpunct}
{\mcitedefaultendpunct}{\mcitedefaultseppunct}\relax
\EndOfBibitem
\bibitem[Bonod \latin{et~al.}()Bonod, Brianceau, and Neauport]{bonod_full-silica_2021}
Bonod,~N.; Brianceau,~P.; Neauport,~J. Full-silica metamaterial wave plate for high-intensity {UV} lasers. \emph{8}, 1372\relax
\mciteBstWouldAddEndPuncttrue
\mciteSetBstMidEndSepPunct{\mcitedefaultmidpunct}
{\mcitedefaultendpunct}{\mcitedefaultseppunct}\relax
\EndOfBibitem
\bibitem[Piccardo \latin{et~al.}()Piccardo, Cernaianu, Palastro, Arefiev, Thaury, Vieira, Froula, and Malka]{piccardo_trends_2025}
Piccardo,~M.; Cernaianu,~M.~O.; Palastro,~J.~P.; Arefiev,~A.; Thaury,~C.; Vieira,~J.; Froula,~D.~H.; Malka,~V. Trends in relativistic laser–matter interaction: the promises of structured light. \emph{12}, 732\relax
\mciteBstWouldAddEndPuncttrue
\mciteSetBstMidEndSepPunct{\mcitedefaultmidpunct}
{\mcitedefaultendpunct}{\mcitedefaultseppunct}\relax
\EndOfBibitem
\bibitem[Dainese \latin{et~al.}()Dainese, Marra, Cassara, Portes, Oh, Yang, Palmieri, Rodrigues, Dorrah, and Capasso]{dainese_shape_2024}
Dainese,~P.; Marra,~L.; Cassara,~D.; Portes,~A.; Oh,~J.; Yang,~J.; Palmieri,~A.; Rodrigues,~J.~R.; Dorrah,~A.~H.; Capasso,~F. Shape optimization for high efficiency metasurfaces: theory and implementation. \emph{13}, 300\relax
\mciteBstWouldAddEndPuncttrue
\mciteSetBstMidEndSepPunct{\mcitedefaultmidpunct}
{\mcitedefaultendpunct}{\mcitedefaultseppunct}\relax
\EndOfBibitem
\bibitem[Lupoiu \latin{et~al.}()Lupoiu, Shao, Dai, Mao, Edee, and Fan]{lupoiu_multi-agentic_2025}
Lupoiu,~R.; Shao,~Y.; Dai,~T.; Mao,~C.; Edee,~K.; Fan,~J.~A. A multi-agentic framework for real-time, autonomous freeform metasurface design. \url{http://arxiv.org/abs/2503.20479}\relax
\mciteBstWouldAddEndPuncttrue
\mciteSetBstMidEndSepPunct{\mcitedefaultmidpunct}
{\mcitedefaultendpunct}{\mcitedefaultseppunct}\relax
\EndOfBibitem
\end{mcitethebibliography}


\providecommand{\latin}[1]{#1}
\makeatletter
\providecommand{\doi}
  {\begingroup\let\do\@makeother\dospecials
  \catcode`\{=1 \catcode`\}=2 \doi@aux}
\providecommand{\doi@aux}[1]{\endgroup\texttt{#1}}
\makeatother
\providecommand*\mcitethebibliography{\thebibliography}
\csname @ifundefined\endcsname{endmcitethebibliography}  {\let\endmcitethebibliography\endthebibliography}{}
\begin{mcitethebibliography}{3}
\providecommand*\natexlab[1]{#1}
\providecommand*\mciteSetBstSublistMode[1]{}
\providecommand*\mciteSetBstMaxWidthForm[2]{}
\providecommand*\mciteBstWouldAddEndPuncttrue
  {\def\EndOfBibitem{\unskip.}}
\providecommand*\mciteBstWouldAddEndPunctfalse
  {\let\EndOfBibitem\relax}
\providecommand*\mciteSetBstMidEndSepPunct[3]{}
\providecommand*\mciteSetBstSublistLabelBeginEnd[3]{}
\providecommand*\EndOfBibitem{}
\mciteSetBstSublistMode{f}
\mciteSetBstMaxWidthForm{subitem}{(\alph{mcitesubitemcount})}
\mciteSetBstSublistLabelBeginEnd
  {\mcitemaxwidthsubitemform\space}
  {\relax}
  {\relax}

\bibitem[Arbabi \latin{et~al.}()Arbabi, Arbabi, Mansouree, Han, Kamali, Horie, and Faraon]{arbabi_increasing_2020}
Arbabi,~A.; Arbabi,~E.; Mansouree,~M.; Han,~S.; Kamali,~S.~M.; Horie,~Y.; Faraon,~A. Increasing efficiency of high numerical aperture metasurfaces using the grating averaging technique. \emph{10}, 7124\relax
\mciteBstWouldAddEndPuncttrue
\mciteSetBstMidEndSepPunct{\mcitedefaultmidpunct}
{\mcitedefaultendpunct}{\mcitedefaultseppunct}\relax
\EndOfBibitem
\bibitem[Khorasaninejad \latin{et~al.}()Khorasaninejad, Zhu, Roques-Carmes, Chen, Oh, Mishra, Devlin, and Capasso]{khorasaninejad_polarization-insensitive_2016}
Khorasaninejad,~M.; Zhu,~A.~Y.; Roques-Carmes,~C.; Chen,~W.~T.; Oh,~J.; Mishra,~I.; Devlin,~R.~C.; Capasso,~F. Polarization-Insensitive Metalenses at Visible Wavelengths. \emph{16}, 7229--7234\relax
\mciteBstWouldAddEndPuncttrue
\mciteSetBstMidEndSepPunct{\mcitedefaultmidpunct}
{\mcitedefaultendpunct}{\mcitedefaultseppunct}\relax
\EndOfBibitem
\end{mcitethebibliography}
\end{document}


\newpage

\section{1. Fill-factor effect on Bloch modes transmission and phase maps}

\begin{figure*}[h!]
  \centering
    \includegraphics[width=0.7\textwidth]{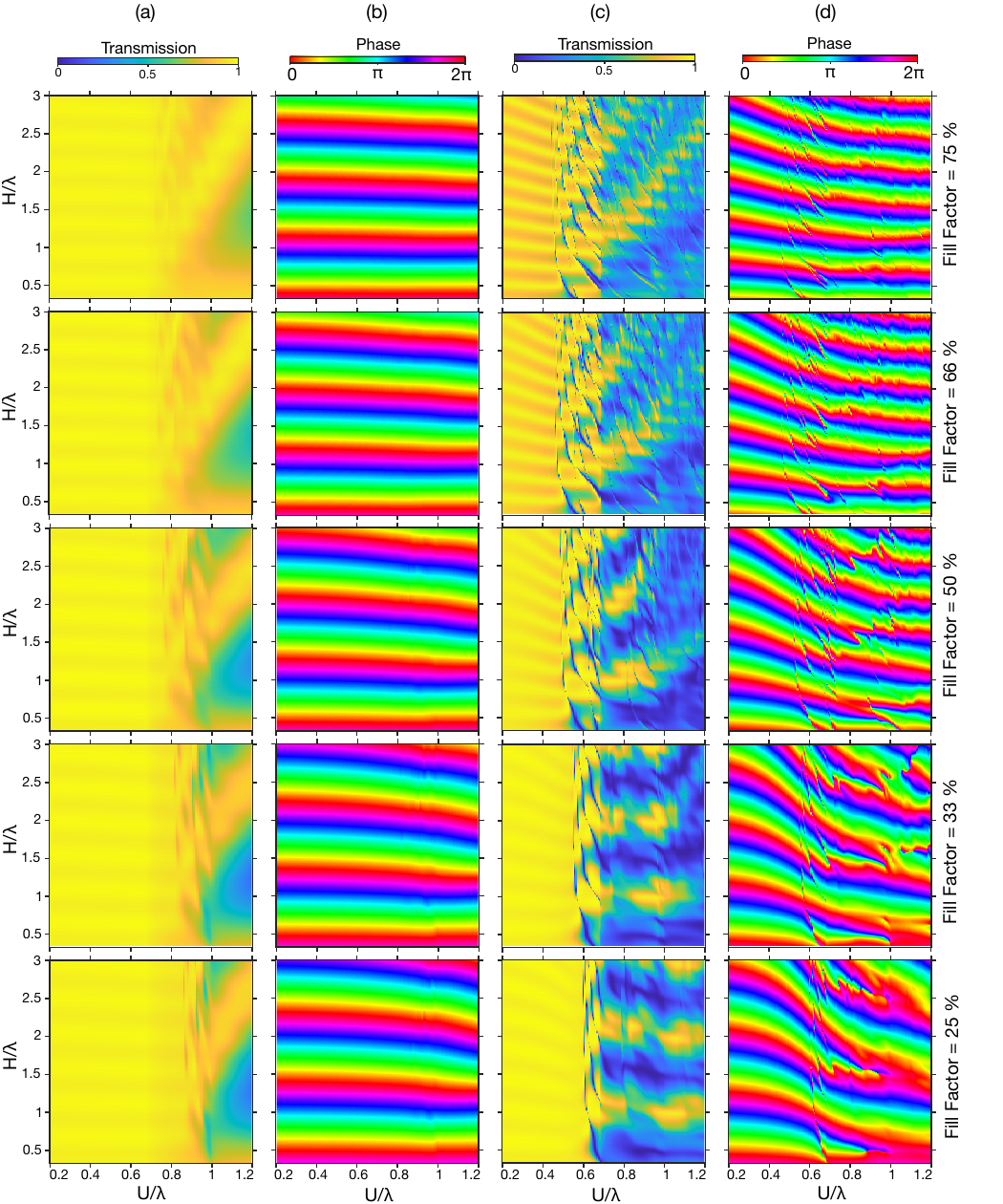}
    \caption{\textbf{Influence of fill factor on modal cut-off.} Color maps of transmission amplitude and phase versus normalized height $H/\lambda$ and unit-cell size $U/\lambda$ for SiO$_2$ \textbf{(a,b)} and TiO$_2$ \textbf{(c,d)} pillars. Rows, from top to bottom, correspond to fill factors of 75\%, 66\%, 50\%, 33\%, and 25\%, respectively. It is clear that in each plot, the region on the right represents areas where multimode interference reduces transmission and disrupts the phase, while in the region on the left, a single Bloch mode enables a high transmittance and smooth phase variance.}
\label{fig:Supplementary_Figure_1}
\end{figure*}

Figure S1 extends the parameter sweep of Figure 2a to multiple fill factors (75\%, 66\%, 50\%, 33\%, and 25\%) for SiO$_2$ and TiO$_2$ pillars, respectively. Across all rows, the boundary separating the single-mode and multimode regimes moves according to the plot in Figure 2b. Left of this boundary, transmission remains high and the phase varies smoothly with $H/\lambda$, consistent with robust single-mode guidance. The fundamental Bloch mode in this region undergoes Fabry–Pérot-like resonances within the pillar. Right of the boundary, the onset of higher-order Bloch modes introduces strong intermodal beating, which, particularly for TiO$_2$, suppresses transmission and scrambles the phase response. These trends confirm that fixing the fill factor at 75\% in Figure 2a captures the essential physics and yields conclusions that are broadly representative over a wide range of geometries.

\newpage

\section*{2. Nanopillar libraries used in simulations of Figure 2}

\begin{figure*}[h!]
  \centering
    \includegraphics[width=0.85\textwidth]{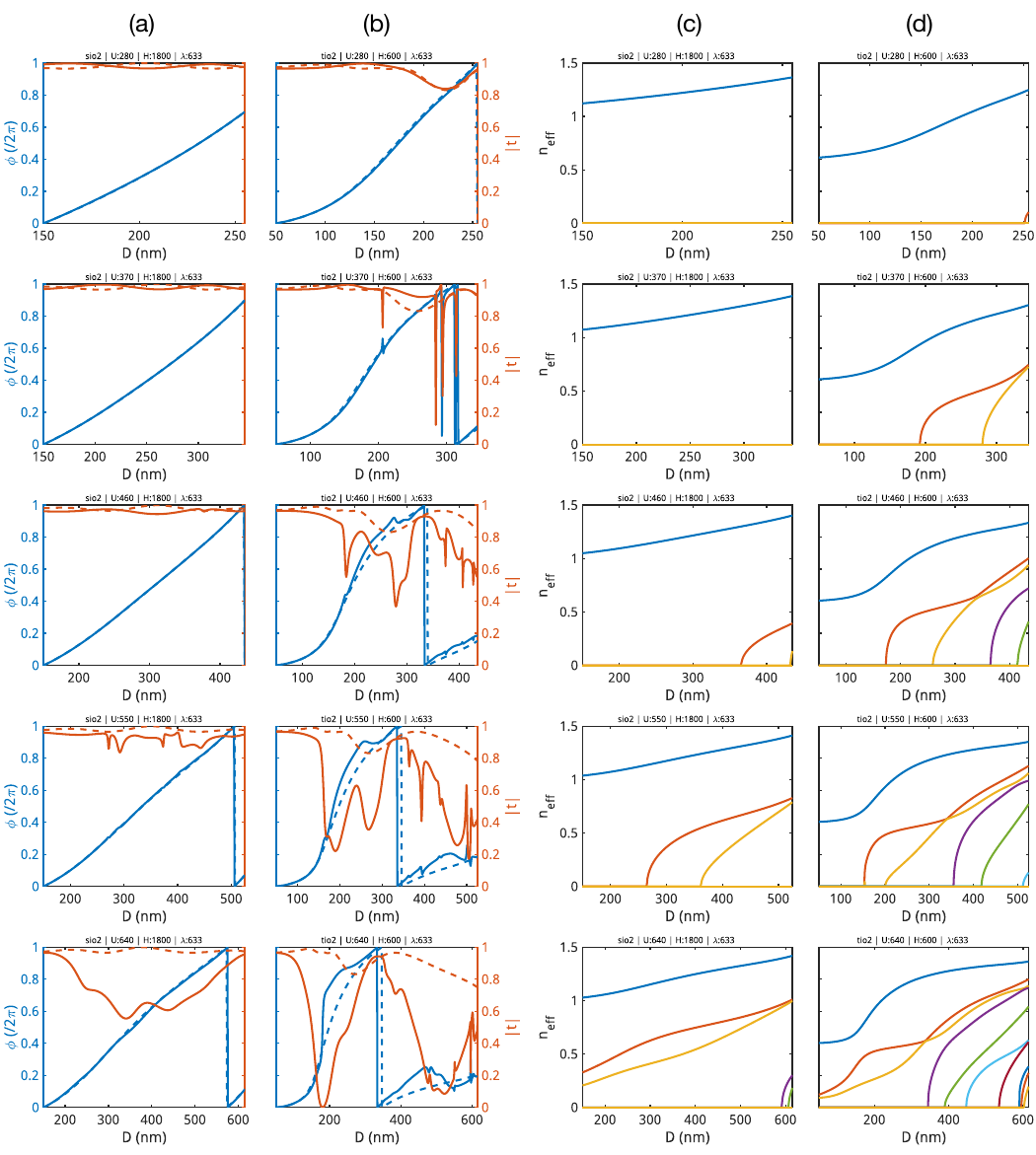}
    \caption{\textbf{Library and Bloch-mode content versus $U/\lambda$} \textbf{(a)} Phase (left axis, $\phi/2\pi$) and transmission amplitude (right axis, $|t|$) for SiO$_2$ pillar libraries as the lateral size $D$ is varied. \textbf{(b)} Phase (left axis, $\phi/2\pi$) and transmission amplitude (right axis, $|t|$) for TiO$_2$ pillar libraries as the lateral size $D$ is varied. The dashed lines in (a) and (b) represent the transmission and phase curves obtained by modelling each nanopillar as a Fabry-Perot cavity whose refractive index is reported in (c) and (d). \textbf{(c)}Effective refractive indices of the Bloch modes supported by the SiO$_2$ pillars in (a). \textbf{(d)}Effective refractive indices of the Bloch modes supported by the TiO$_2$ pillars in (b).}
    \label{fig:Supplementary_Figure_5}
\end{figure*}

Here we show some of the generated pillar (square shape) libraries used for the simulations in Figure 2 of the main text. Each library is computed under periodic boundary conditions (local periodic approximation) at an aspect ratio AR $\leq$ 12 (height/width of the nanopillar). The height is dictated then by the smallest dimensions in the library, which for TiO$_2$ is $H = 12 \times50\ nm = 600\ nm$ and for SiO$_2$ is $H=12\times150\ nm = 1800\ nm$.

Rows in Figure S2 correspond to $U=280,\,370,\,460,\,550,\,640$ nm at $\lambda_d=632$ nm ($U/\lambda = 0.44,\, 0.58,\, 0.73,\, 0.87,\, 1.01$).

Columns (a,b) report the simulated transmission amplitude $|t|$ (right axis, solid red line) and phase $\phi$ (left axis, solid blue line) versus pillar diameter $D$. The blue dashed lines represent the phase calculated by $\phi = 2\pi n_{\text{eff}} H/\lambda_d$, where $H$ is the height of the pillar, and $n_{\text{eff}}$ is the effective refractive index of the fundamental Bloch mode. The effective refractive index is determined using Bloch wave theory, where the electromagnetic fields are expanded in a Fourier basis along the $x$ and $y$ directions, and the pillar is considered infinite in $z$. The red dashed line represents the transmission calculated by modeling each meta-atom as a Fabry-Perot cavity whose refractive index is given by the $n_{\text{eff}}$ of the propagating fundamental Bloch mode. Columns (c,d) plot the effective indices of the Bloch modes supported by the corresponding pillars as $D$ increases, thereby revealing when additional guided modes appear.  

For $U/\lambda=0.44$ in the TiO$_2$ case, column (b) shows a near-linear phase sweep with $L$ and a transmission amplitude close to unity. The transmission and phase curves are extremely well approximated if the nanostructures are modeled by the Fabry-Perot model. Column (d) confirms single-mode guidance over most of the $D$ range, in agreement with the cut-off in Figure 2b. As $U/\lambda$ is increased to $0.58$, an additional mode enters, and, for even larger $U/\lambda,$ panel (d) reports multiple curves relative to multiple modes, and their beating produces pronounced amplitude ripples and non-linear phase variations in panel (a). In this scenario, the Fabry-Perot model does not accurately predict the transmission curve.

The SiO$_2$ libraries in column (a) remain smooth and with high transmission up to $U/\lambda = 0.73$. The Fabry-Perot model correctly predicts the phase and transmission curves. Column (c) indicates that the structure stays single mode for much larger $D$ and $U/\lambda$; even when additional modes start to appear near the upper end of the scan, their impact on $|t|$ and $\phi$ is minor and the phase progression with $D$ remains nearly linear. 

Figure S2 shows that the onset of multiple guided modes is strictly related to the library transmission and phase curve, and because the structural cut-off occurs at much larger $U/\lambda$ for SiO$_2$ (cf. Figure 2), low-index pillars preserve a clean, monotonic phase law and high $|t|$ over a broader design space. For AR$=12$, TiO$_2$ libraries are reliable at small $U/\lambda$ (subwavelength and well below cut-off), whereas SiO$_2$ libraries remain robust even as $U/\lambda$ approaches and exceeds unity.

\newpage

\section*{3. Effect of aspect ratio on high- and low-index meta-gratings}

\begin{figure*}[h!]
  \centering
    \includegraphics[width=0.65\textwidth]{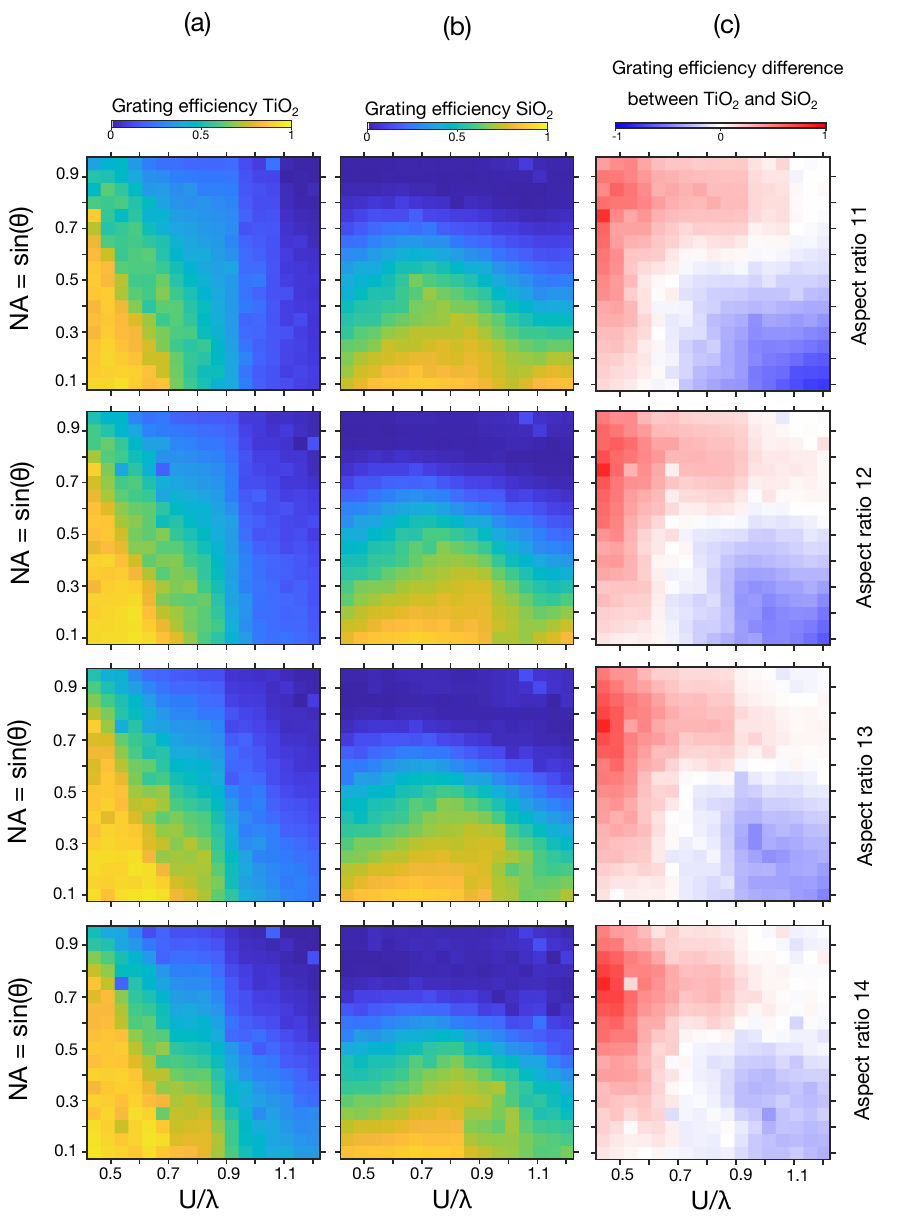}
    \caption{\textbf{Aspect-ratio dependence of grating efficiency.} \textbf{(a,b)} First-order diffraction efficiency $\eta$ versus normalized period $U/\lambda$ and numerical aperture NA = $\sin(\theta)$ for TiO$_2$ \textbf{(a)} and SiO$_2$ \textbf{(b)} gratings. Rows correspond to aspect ratios (AR = height/width) of 11, 12, 13, and 14. \textbf{(c)} Difference maps $\Delta\eta=\eta_{\mathrm{TiO}_2}-\eta_{\mathrm{SiO}_2}$ with red (blue) indicating TiO$_2$ (SiO$_2$) superiority. As AR decreases, the blue region becomes more pronounced because high-index meta-atoms must broaden laterally to meet the $(0,2\pi)$ phase coverage, crossing the structural cut-off and exciting multiple Bloch modes that lower transmission and grating efficiency. At larger AR, the opposite is true and the red region becomes more pronounced at the expense of the blue region.}
    \label{fig:Supplementary_Figure_2}
\end{figure*}

To assess how thickness constraints influence the material choice, we repeated the $(U/\lambda,\ \mathrm{NA})$ sweep used in Figure 2c while varying the aspect ratio (AR) of the meta-atoms. The height is dictated then by the smallest dimensions in the library, which for TiO$_2$ is $H = AR\times50\ nm$ and for SiO$_2$ is $H=AR\times150\ nm$. Figure S2 reports first-order diffraction efficiency $\eta$ for TiO$_2$ and SiO$_2$ gratings together with the difference map $\Delta\eta=\eta_{\mathrm{TiO}_2}-\eta_{\mathrm{SiO}_2}$ in panel (a), (b) and (c) respectively for different AR values ranging from 11 to 14. These maps complement Figure 2c by revealing how thickness rebalances the competition between modal content within each pillar and interpillar coupling across the lattice.

Two robust features persist for all ARs. First, at small $U/\lambda$ and modest NA, TiO$_2$ remains highly efficient owing to its strong field confinement, which suppresses lateral leakage and keeps the locally periodic approximation (LPA) accurate. Second, as $U/\lambda$ increases, TiO$_2$ quickly encounters its structural cut-off (\textit{cf}. Figure 2b), beyond which most of the pillars in the library are multi-mode, resulting in a collapse of the efficiency. By contrast, SiO$_2$ libraries stay single-mode over a much larger $U/\lambda$ range and thus retain high transmission, resulting in greater grating efficiency.
Changing AR primarily shifts how each material satisfies the required phase ramp for a given $(U/\lambda,\ \mathrm{NA})$. Lower AR forces the use of wider meta-atoms (for the same available height) to accumulate the necessary phase. This lateral broadening pushes more and more high-index TiO$_2$ structures into their multimode regime, reducing the overall transmission and efficiency of the grating. SiO$_2$, having a much higher single-mode threshold, tolerates this increase in the broadening of the pillar diameter without triggering additional guided modes.

From AR = 14 to AR = 11, the blue area in the difference maps (where $\Delta\eta<0$ and SiO$_2$ is superior) slowly expands toward smaller $U/\lambda$ and higher NA. Physically, reducing AR increases the transverse dimension of the TiO$_2$ pillars required to meet the phase demand, thereby exciting higher-order modes and reducing transmission. In parallel, the same reduction in AR increases intercoupling for both materials.

Taken together, Figure S3 indicates that thickness-limited implementations (smaller AR) systematically enlarge the parameter region in which SiO$_2$ outperforms TiO$_2$, extending even into portions of the high-NA, subwavelength-lattice domain. Conversely, when tall pillars are feasible (larger AR) and minimum feature sizes allow small $U/\lambda$, TiO$_2$ maintains an advantage through the ability to operate in a single-mode regime. Thus, the optimal material is set jointly by $(U/\lambda,\ \mathrm{NA})$ \emph{and} the available AR: decreasing AR shifts the balance toward SiO$_2$; increasing AR shifts it toward TiO$_2$.

\newpage

\section*{4. Metagrating design choice}

\begin{figure*}[h!]
  \centering
    \includegraphics[width=0.9\textwidth]{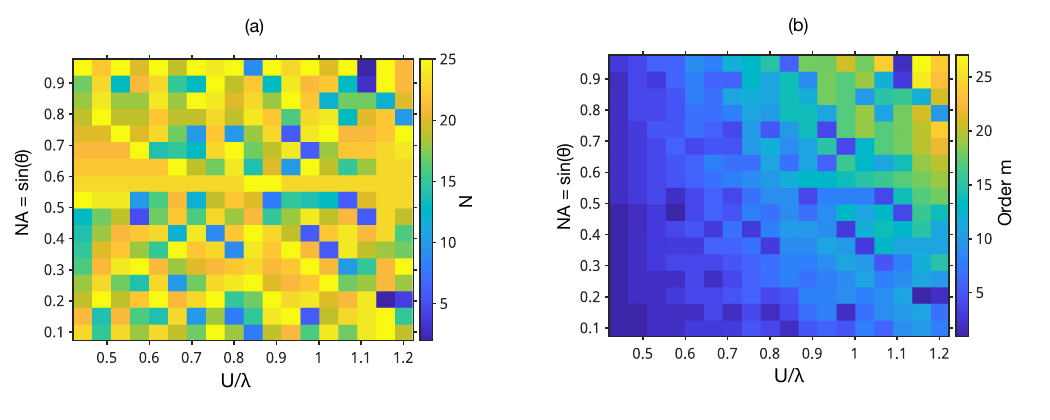}
    \caption{\textbf{Selection rule used to generate and simulate gratings.} For each $(U/\lambda,\ \mathrm{NA})$, we searched integers $N \le 25$ such that $\tilde m=N\,U\sin\theta/\lambda$ stays within $0.05$ of an integer. \textbf{(a)} Number of pillars $N$ used for each $(U/\lambda,\ \mathrm{NA})$.  \textbf{(b)} Diffraction order $m$ associated with each simulated grating.}
\label{fig:Supplementary_Figure_3}
\end{figure*}

To generate the $(U/\lambda,\ \mathrm{NA})$ design maps shown in Figure 2, we followed the procedure detailed below. 
For each case, we selected the largest possible number of pillars, $N \leq 25$, such that the grating order
\begin{equation}
    \tilde m = \frac{N U \sin \theta}{\lambda}
\end{equation}
fell within a small tolerance of an integer. Specifically, we required
\begin{equation}
    \epsilon = \bigl|\tilde m - \mathrm{round}(\tilde m)\bigr| < 0.05.
\end{equation}

\noindent When this condition is satisfied, the grating period $N U$ obeys the scalar grating equation
\begin{equation}
    m \lambda = N U \sin \theta, \quad m = \mathrm{round}(\tilde m).
\end{equation}
This ensures that the imposed phase ramp repeats almost exactly an integer number of times within one grating period.  In turn, this averages out the ambiguity introduced by a global constant phase offset across the lattice, which would otherwise redistribute power among neighboring diffraction orders---a problem that becomes especially severe at high numerical apertures \cite{arbabi_increasing_2020}.

For every set $(U,\mathrm{NA}, N,\tilde m)$, we simulated the corresponding grating built up by $N$ unit cells $U$ and read out the efficiency in the diffraction order $m=\mathrm{round}(\tilde m)$. Panels (a) and (b) in Figure S4 show the integer $N$ and the corresponding order $m$ found for each $(U,\mathrm{NA})$.

\newpage

\section*{5. Comparison of TiO$_2$ and SiO$_2$ bulk material dispersions}

\begin{figure*}[h]
  \centering
    \includegraphics[width=0.5\textwidth]{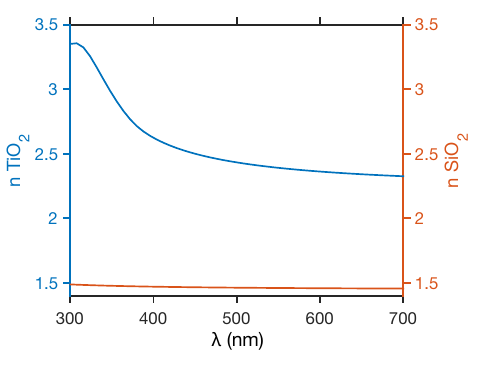}
    \caption{Bulk refractive index $n_{\text{bulk}}$ of TiO\textsubscript{2} and SiO\textsubscript{2} as a function of wavelength.}
    \label{fig:Supplementary_Figure_4}
\end{figure*}

Supplementary Figure S5 compares the wavelength dependence of the bulk refractive indices $n(\lambda)$ of TiO$_2$ and SiO$_2$ across the visible band measured with ellipsometry. As is well known and tabulated (\textit{e.g.}, Sellmeier/Palik data), TiO$_2$ exhibits a pronounced reduction of the refractive index $n$ with wavelength $\lambda$, whereas fused silica is nearly dispersionless over the same range. This difference in \emph{material} dispersion complements the modal picture developed in the main text.

\newpage

\section*{6. Nanopillar libraries used in simulations for Figure 3}

\begin{figure*}[h!]
  \centering
    \includegraphics[width=0.7\textwidth]{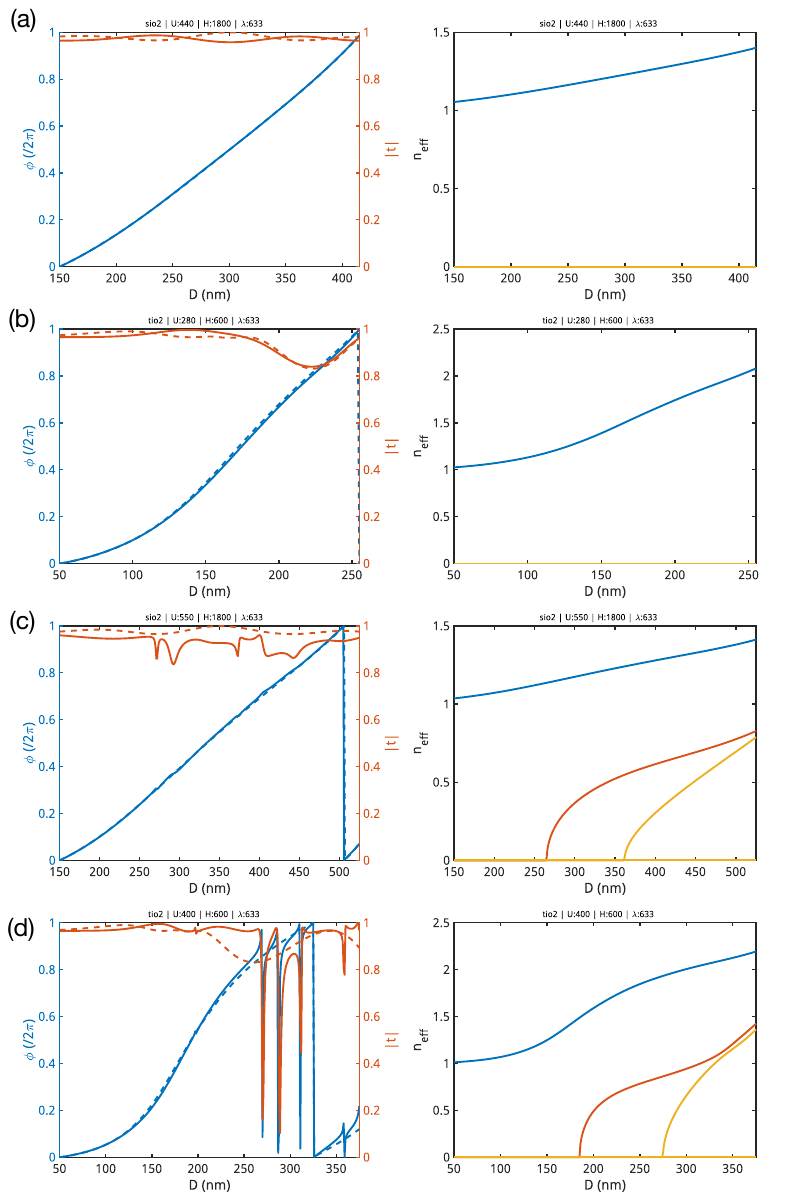}
    \caption{\textbf{Library and Bloch-mode content versus $U/\lambda$} \textbf{(a)} The right plot shows the phase (left axis, $\phi/2\pi$) and transmission amplitude (right axis, $|t|$) for the SiO$_2$ pillar library at $U=440\ nm$. The left plot shows the effective refractive indices of the Bloch modes supported by the SiO$_2$ pillars. \textbf{(b)} Same as (a) but for TiO$_2$ pillar library at $U=440\ nm$. \textbf{(c)} Same as (a) but for SiO$_2$ pillar library at $U=550\ nm$. \textbf{(d)}Same as (a) but for TiO$_2$ pillar library at $U=400\ nm$..}
    \label{fig:Supplementary_Figure_5}
\end{figure*}

Simulated libraries used to generate the gratings simulated in Figure 3. In Figures S6a and S6b, the left plots show the phase and transmission of SiO$_2$ and TiO$_2$, respectively, when the library is fully single-mode. In Figures S6c and S6d, the left plots show the phase and transmission of SiO$_2$ and TiO$_2$, respectively, when the library is multi-mode. For the TiO$_2$ libraries, the phase ramp and the transmission are more evident, negatively affecting the grating efficiencies in Figure 3d.

The libraries are built with aspect ratio AR $\leq$ 12 (height/width of the nanopillar). The height is dictated then by the smallest dimensions in the library, which for TiO$_2$ is $H = 12 \times50\ nm = 600\ nm$ and for SiO$_2$ is $H=12\times150\ nm = 1800\ nm$.
\newpage

\section{7. Fabrication method}

\begin{figure*}[h!]
  \centering
    \renewcommand{\thefigure}{S7}
    \includegraphics[width=0.72\textwidth]{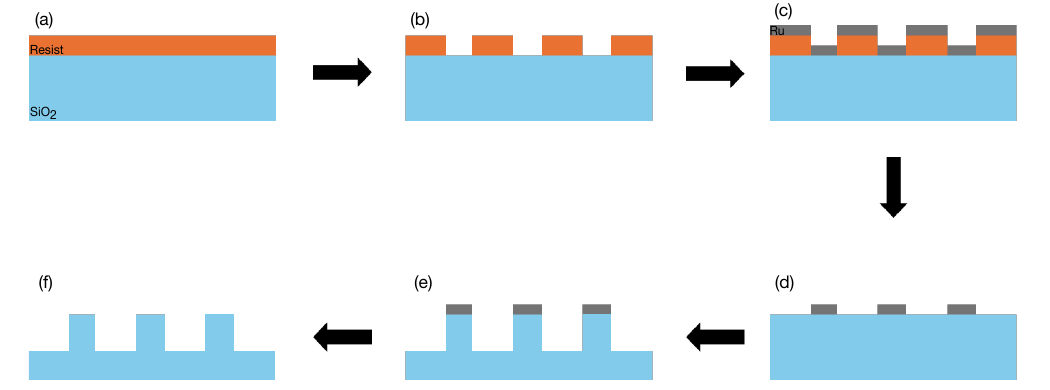}
    \caption{\textbf{Stepwise illustration of the metasurface fabrication.} \textbf{(a)} Spin-coating of PMMA resist (300 nm thick) on a 500 $\mu m$ UV-grade fused silica substrate. \textbf{(b)} Electron-beam lithography followed by resist development. \textbf{(c)} Deposition of a 50 nm Ru hard mask by magnetron sputtering. \textbf{(d)} Lift-off in acetone produces the patterned Ru mask. \textbf{(e)} Transfer of the pattern into the silica using ICP-RIE. \textbf{(f)} Final structure after selective Ru removal, revealing monolithic silica nanopillars.}
    \label{fig:Supplementary_Figure_7}
\end{figure*}

We begin with 500 $\mu m$ thick, UV-grade, double-side polished fused silica wafers.  A positive double layer of e-beam resist (PMMA 950K, PMMA 495K) is spin-coated, resulting in a $\sim$300 nm resist layer (Figure S7a). The metasurface patterning is carried out using electron-beam lithography (Elionix-150 system operated at 150 kV). After exposure, the resist is developed in MIBK:IPA 1:3 solution (Figure S7b). A 50 nm ruthenium (Ru) layer is subsequently deposited across the entire substrate by magnetron sputtering (Figure S7c). Lift-off in acetone yields the patterned Ru hard mask (Figure S7d). Pattern transfer into the silica is achieved using inductively coupled plasma reactive ion etching (ICP-RIE, SPTS Omega 2CL) (Figure S7e). Finally, the residual Ru mask is stripped with a dedicated metal-etch process in the same ICP tool, revealing the final nanopillar array (Figure S7f).

\newpage

\section{8. Experimental setup}

\begin{figure*}[h!]
  \centering
      \renewcommand{\thefigure}{S8}
    \includegraphics[width=0.9\textwidth]{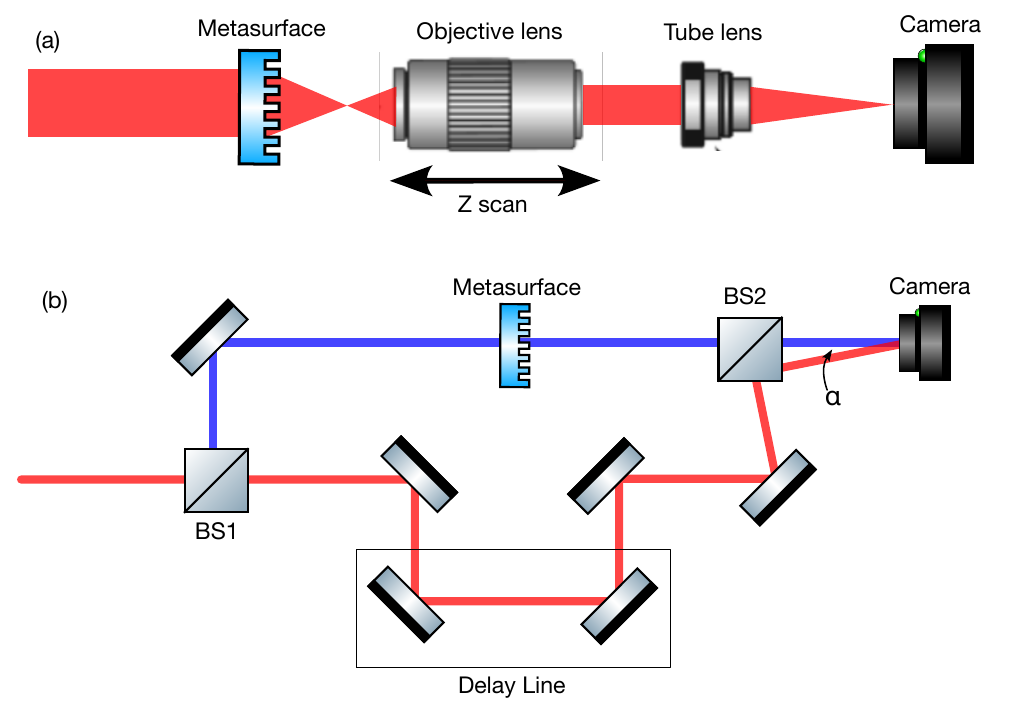}
    \caption{ \textbf{Experimental Setups.} \textbf{(a)} Schematic of the microscope setup for metalens characterization. A collimated laser beam (Thorlabs RC12APC-P01) illuminates the metasurface. The focused field is imaged by a Mitutoyo objective (NA = 0.55) and a tube lens onto a camera (Basler daA3840-45um). \textbf{(b)} Off‑axis digital holography setup for OAM phase retrieval. The beam is split at BS1 into sample (through the metasurface) and reference arms; the reference path includes a delay line to match optical path lengths. At BS2, the two fields recombine with a small tilt $\alpha$, generating an interferogram on the camera. The complex field is obtained by Fourier‑domain filtering of the off‑axis hologram, yielding the amplitude and the helical phase profile.}
    \label{fig:Supplementary_Figure_8}
\end{figure*}

The optical response of the metalenses fabricated was characterized with a home-built microscope setup shown in Figure S8a. A collimated (Thorlabs RC12APC-P01) laser beam is directed onto the metasurface, focusing the incident light into a diffraction-limited spot. The focal field distribution was collected using a Mitutoyo objective lens (NA=0.55) and relayed to a monochrome camera (Basler daA3840-45um) using a tube lens. Axial scans were performed by translating the objective in 1 $\mu$m increments with a motorized stage, allowing reconstruction of the beam evolution along the propagation direction. The acquired intensity maps were stacked to generate focal profiles. 

To retrieve the OAM helical phase, an off‑axis digital holography interferometer shown in Figure S8b is implemented. The collimated laser is split using a beamsplitter (BS1) into a sample arm, which passes through the metasurface, and a reference arm that bypasses it. The reference arm is routed through a free‑space delay line to equalize the optical path and ensure temporal coherence at the detector. The two fields are recombined at a second beamsplitter (BS2) with a small relative angle $\alpha$, producing spatial carrier fringes on the camera. The complex field $E(x,y)$ is reconstructed from a single interferogram by Fourier filtering: the first off‑axis sideband is isolated, shifted to the origin, and inverse‑transformed to recover amplitude and phase. The retrieved wrapped phase directly reveals the OAM topological charge and phase singularity structure shown in the Main text.

To measure the absolute diffraction efficiency, we placed an iris upstream of the metasurface and adjusted it so that the laser illuminated only the patterned area. For gratings, a power meter was positioned to record the diffracted power in the specific order of interest. For lenses, the power meter was placed at the focus of the tube lens to capture the power of the laser focused. Reference measurements were taken with the iris set to the same aperture while the metasurface was shifted out of the beam, so the meter measured only the power transmitted through the substrate.

To calculate the Strehl ratios of metalenses \cite{khorasaninejad_polarization-insensitive_2016}, the measured integrated powers of PSFs are normalized to those of ideal Airy pattern functions up to the third zero of the Airy function.

\newpage

\section{9. Achromatic and chromatic criteria}

In this section, we analyze under what conditions the achromatic response of a single meta-atom translates into an efficient chromatic or achromatic response of the entire metasurface. It is important to distinguish between the broadband behavior of individual meta-elements and that of the metasurface as a whole. Let us denote the target phase profile of the device as
$$
\phi(x,y,\lambda_d, p_d),
$$
where $\lambda_d$ is the design wavelength and $p_d$ is a parameter encoding the desired optical functionality. For example:
\begin{itemize}
    \item For a linear grating with phase profile
    $$
    \phi(x,y,\lambda_d,\theta_d) = \frac{2\pi}{\lambda_d} \, x \sin\theta_d,
    $$
    the parameter \(p_d\) corresponds to the deflection angle \(\theta_d\).
    
    \item For a hyperbolic lens with
    $$
    \phi(r,\varphi,\lambda_d,f_d) = \frac{2\pi}{\lambda_d}\left(\sqrt{r^2+f_d^2}-f_d\right),
    $$
    $p_d$ corresponds to the focal length $f_d$.
    \item For a parabolic lens with
    $$
    \phi(r,\varphi,\lambda_d,f_d) = -\frac{2\pi r^2}{\lambda_d f_d},
    $$
    $p_d$ corresponds to the focal length $f_d$.    
    \item For an OAM phase plate with
    $$
    \phi(r,\varphi,\lambda_d,\ell) = \ell \varphi,
    $$
    $p_d$ corresponds to the topological charge $\ell$.
\end{itemize}

When the target phase profile is \emph{independent} of wavelength, i.e. $\phi(x,y,\lambda_d,p_d) = \phi(x,y,p_d)$, as in the case of OAM beams or holograms, the achromatic response of the individual elements naturally leads to an achromatic metasurface. Each element imparts the same phase shift at every wavelength, so the global phase profile is preserved across the spectrum. This behavior is confirmed experimentally in Fig.~4b, where the same OAM phase plate works consistently at 632 nm and 532 nm.  

In contrast, when the desired phase profile \emph{depends} on wavelength, as in gratings or lenses, an achromatic response at the pillar level does not automatically guarantee broadband operation of the full device. The device can only realize the required optical functionality across different wavelengths, with dispersive and chromatic behavior, provided that the phase profile equation can be inverted to express the functionality parameter $p$ as a function of $\lambda$, $\lambda_d$, and $p_d$.
 The dispersion of the functionality $p(\lambda)$ comes directly from the dependence of the phase profile on wavelength. For example, a metagrating whose elements impose a linearly increasing phase ramp across a broadband spectrum---owing to their individual achromatic response---will deflect light efficiently; however, different wavelengths will be deflected to different angles. Thus, the grating as a whole is not achromatic, such that it does not deflect all wavelengths to the same angle. This can be seen directly from the equation of the phase profile: 

$$\phi(x,\lambda_d,\theta_d) = \frac{2\pi}{\lambda_d} \, x \sin(\theta_d) = \frac{2\pi}{\lambda} \, x \sin(\theta) \quad \forall x $$

which gives the dispersion relation of the deflecting angle 
$$ \theta(\lambda) = \arcsin\left(\lambda \frac{\sin(\theta_d)}{\lambda_d}\right).$$
The same can be found for a parabolic lens. When each element is achromatic, then 

$$\phi(r,\varphi,\lambda_d,f_d) = -\frac{2\pi r^2}{\lambda_d f_d} = -\frac{2\pi r^2}{\lambda f} \quad \forall r $$
and the dispersion relation of the focal spot is 
$$f(\lambda) = \frac{\lambda_d f_d}{\lambda}.$$

On the other hand, it is not possible to invert the hyperbolic phase profile to have an expression for the dispersion of the focal length. In this case, then, having each element behave achromatically does not result in an efficient chromatic behavior of the device, and the optical functionality breaks at wavelengths different from the design wavelength.  

One can summarize the previous concepts in the following criteria. A device behaves achromatically whenever 

\begin{equation}\label{eq:achro}
    \phi(x,y,\lambda_d,p_d) = \phi(x,y,\lambda,p_d) \quad \forall x,y
\end{equation}
is satisfied for a given set of wavelengths $\lambda\neq \lambda_d$. Conversely, a device behaves chromatically when 

\begin{equation}\label{eq:chro}
    \phi(x,y,\lambda_d,p_d) = \phi(x,y,\lambda,p) \quad \forall x,y
\end{equation}
is satisfied for a given set of wavelengths $\lambda\neq \lambda_d$ and there exists an optical functionality parameter $p\neq p_d$, that may depend on the remaining parameters $\lambda$, $\lambda_d$, and $p_d$.

\newpage

\section{10. Broadband measurement of hyperbolic phase metalenses with NA = 0.1 and NA = 0.5}

\begin{figure*}[h!]
  \centering
   \renewcommand{\thefigure}{S10}
    \includegraphics[width=0.8\textwidth]{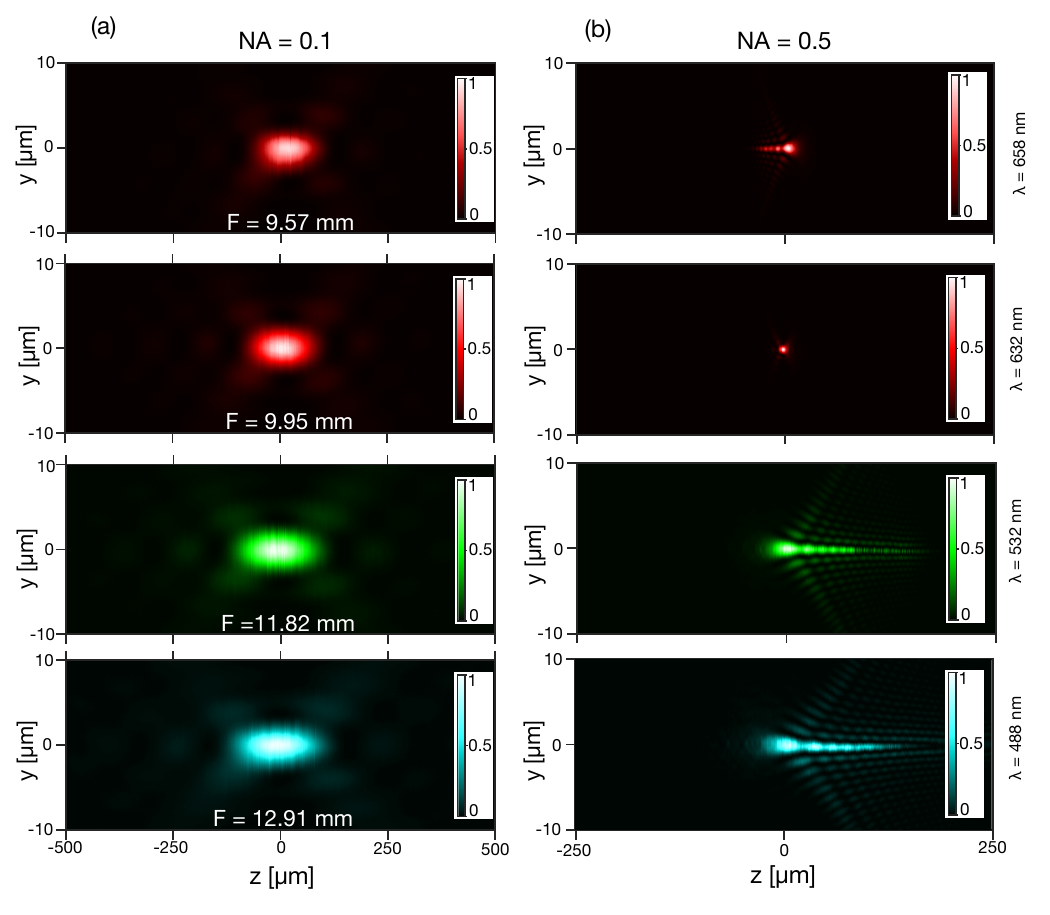}
    \caption{Focusing profile for a fabricated silica lens of \textbf{(a)} NA = 0.1 and \textbf{(b)} NA = 0.5 at various wavelengths. Colorbar represents the intensity of the image. From top to bottom, the excitation wavelengths are: 658 nm, 632 nm, 532 nm, and 488 nm. The aberration behavior is mainly due to the breaking of the criteria explained in Supplementary Note 6.}
    \label{fig:Supplementary_Figure_10}
\end{figure*}

Two hyperbolic silica-metalenses were fabricated with numerical apertures NA = 0.1 and NA = 0.5. The focusing behavior of the metalenses was quantified by axial ($z$) scans at multiple wavelengths for NA = 0.1 in Figure S10a and NA = 0.5 in Figure S10b. For the low numerical aperture lens, a diffraction-limited (see Figure 4 in the main text) focal spot is observed at each wavelength, with focal planes located at $f=9.04$ mm (658 nm), $f=9.95$ mm (632 nm), $f=11.07$ mm (532 nm), and $f=12.45$ mm (488 nm), evidencing a smooth, monotonic chromatic shift of the focus that follows the scaling $f(\lambda) = \frac{\lambda_d f_d}{\lambda}$. In fact, for low numerical aperture, it is possible to approximate the phase profile of the hyperbolic lens via a Taylor expansion in the radial coordinate, recovering the parabolic phase profile. The scaling of $f(\lambda)$ then comes from the achromatic behavior of the silica nanopillars, as discussed in section S9. On the other hand, one cannot perform a Taylor expansion for a high numerical aperture hyperbolic lens, and the achromatic behavior of each single nanopillar does not lead to the same optical functionalities at different wavelengths. It can be seen in Figure S10b that at wavelengths differing from the design wavelength $\lambda_d = 632$ nm, a strong chromatic aberration deteriorates the focusing profile of the device. 

\newpage

\section*{11. Metalens imaging}

\begin{figure*}[h!]
    \centering
    \renewcommand{\thefigure}{S11}
    \includegraphics[width=0.8\textwidth]{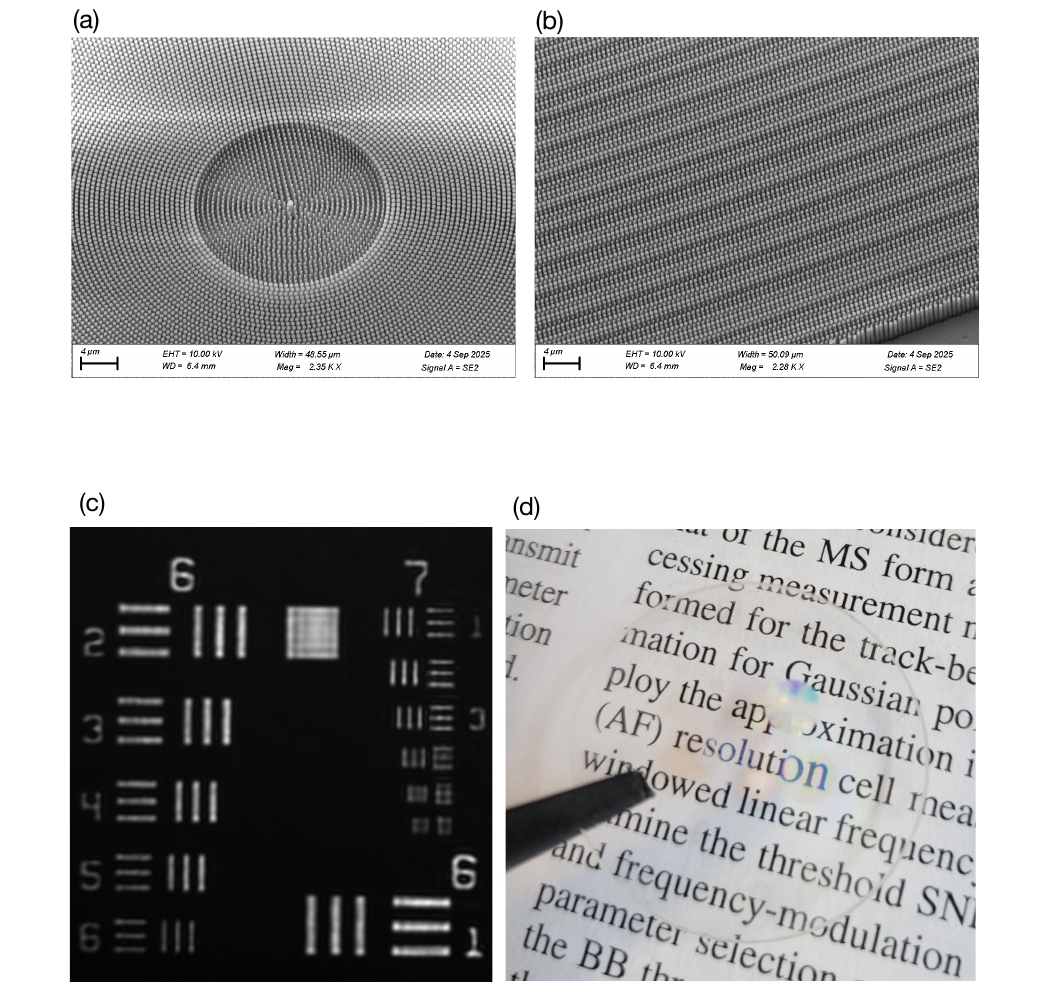}
        \caption{\textbf{Silica metalens fabrication and characterization.} \textbf{(a,b)} SEM of the metalens showing the concentric phase layout and locally periodic nanopillar array.  \textbf{(c)} Image of a 1951 USAF resolution chart formed by the silica metalens at 632 nm (NA = 0.1). \textbf{(d)} Photograph of a 2 mm‑aperture silica metalens magnifying printed text under white‑light illumination.}
        \label{fig:Supplementary_Figure_11}
\end{figure*}

Figures S11a–b display Scanning Electron Microscope (SEM) images of the fabricated silica metalens, showing the concentric phase profile and the uniform nanopillar lattice. Imaging tests further confirm proper lens operation: at the design wavelength (632 nm), the NA = 0.1 lens forms a high-contrast image of a 1951 USAF resolution target (Figure S11c). A 2 mm silica lens produces clear magnification of printed text under broadband illumination (Figure S11d).

\newpage

\section*{12. SEM characterization of the silica vortex-generating metasurface}

\begin{figure*}[h]
  \centering
      \renewcommand{\thefigure}{S12}
    \includegraphics[width=0.8\textwidth]{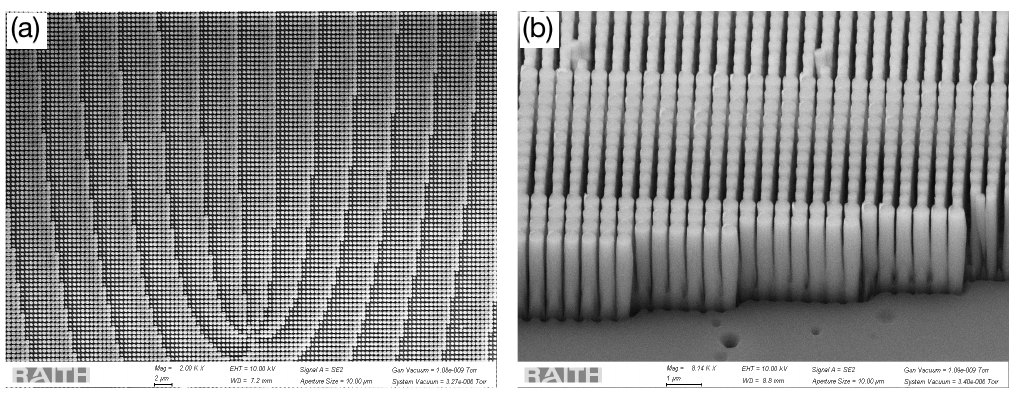}
    \caption{\textbf{Silica vortex generator metasurface characterization} \textbf{(a)} Top-view SEM image showing a large, defect-free region with a uniform lattice and a smooth spatial variation of the pillar diameter. \textbf{(a)} SEM highlighting vertical nanopillars etched in fused silica.}
 \label{fig:Supplementary_Figure_12}
\end{figure*}

Figure S12a shows a representative top view: the locally varying pillar diameters encode the target phase profile to generate an OAM beam with topological charge $\ell=5$ with uniform pitch over a large field, and no stitching artifacts or pillar collapse are observed.  Figure S12b confirms that the pillars are fully etched into the silica substrate and remain mechanically robust.

\bibliography{references}